\documentclass[sigconf, nonacm]{acmart}

\usepackage[noend]{algorithmic}
\usepackage{array,multirow}
\usepackage{xspace}
\usepackage{pifont}
\usepackage{adjustbox}
\usepackage{xcolor}
\usepackage{pbox}
\usepackage{graphicx}
\usepackage{subfigure}
\usepackage{colortbl}

\def\ojoin{\setbox0=\hbox{$\bowtie$}%
	\rule[0.15ex]{.22em}{.6pt}\llap{\rule[0.9ex]{.22em}{.6pt}}}

\def\fullouterjoin{\mathbin{\ojoin\mkern-5.5mu\bowtie\mkern-5.5mu\ojoin}}

\definecolor{mygrey}{RGB}{230,230,240}
\definecolor{myblue}{RGB}{175, 238, 235}
\definecolor{mywt}{RGB}{255, 230, 153}
\definecolor{mytp}{RGB}{169, 209, 142}
\definecolor{mywh}{RGB}{239, 182, 176}
\definecolor{mywf}{RGB}{180, 199, 231}

\newcommand\vldbdoi{10.14778/3461535.3461539}
\newcommand\vldbpages{1489 - 1502}
\newcommand\vldbvolume{14}
\newcommand\vldbissue{9}
\newcommand\vldbyear{2021}
\newcommand\vldbauthors{Rong Zhu, Ziniu Wu, Yuxing Han, Kai Zeng, Andreas Pfadler, Zhengping Qian, Jingren Zhou, Bin Cui}
\newcommand\vldbtitle{FLAT: Fast, Lightweight and Accurate Method for Cardinality Estimation} 

\newcommand\vldbpagestyle{empty}

\newcommand{\CE}{\textsf{CardEst }}
\newcommand{\CEend}{\textsf{CardEst}}

\newcommand{\Card}{\textsf{Card}}

\newcommand{\pa}{\text{pa}}

\newcommand{\FSPN}{\textsf{FSPN }}
\newcommand{\FSPNend}{\textsf{FSPN}}
\newcommand{\FLAT}{\textsf{FLAT }}
\newcommand{\FLATend}{\textsf{FLAT}}

\newcommand{\T}{\textit{\textsc{t}}}

\newcommand{\N}{\textit{\textsc{n}}}
\newcommand{\A}{\textit{\textsc{a}}}
\newcommand{\B}{\textit{\textsc{b}}}

\newcommand{\E}{\textit{\textsc{e}}}

\renewcommand{\C}{\textit{\textsc{c}}}
\renewcommand{\L}{\textit{\textsc{l}}}
\newcommand{\R}{\textit{\textsc{r}}}

\newcommand{\F}{\mathcal{F}}

\newcommand{\myskip}{\vspace{0em}}

\newcommand{\revise}[1]{\textcolor{black}{#1}}

\allowdisplaybreaks[4] 

\begin{document}

\title{\textsc{FLAT}: Fast, Lightweight and Accurate Method \break
	for Cardinality Estimation}

\author{Rong Zhu$^{1, \#}$, Ziniu Wu$^{1, \#}$,  Yuxing Han$^{1}$, Kai Zeng$^{1, *}$, Andreas Pfadler$^{1}$, \break Zhengping Qian$^{1}$, Jingren Zhou$^{1}$, Bin Cui$^{2}$}
\affiliation{
	\smallskip \smallskip
	\institution{\LARGE \textit{$^{1}$Alibaba Group, $^{2}$Peking University}}
	\smallskip \smallskip
	\state{\Large \textsf{$^1\{$red.zr, ziniu.wzn, yuxing.hyx, zengkai.zk, andreaswernerrober, zhengping.qzp, \break jingren.zhou$\}$@alibaba-inc.com
			$^2$bin.cui@pku.edu.cn
}}}

\begin{abstract}
Query optimizers rely on accurate cardinality estimation (\CEend) to produce good execution plans. The core problem of \CE is how to model the rich joint distribution of attributes in an accurate and compact manner. Despite decades of research, existing methods either over-simplify the models only using independent factorization which leads to inaccurate estimates, or over-complicate them by lossless conditional factorization without any independent assumption which results in slow probability computation. In this paper, we propose \FLATend, a \CE method that is simultaneously \emph{\underline{f}ast} in probability computation, \emph{\underline{l}ightweight} in model size and 
\emph{\underline{a}ccurate} in es\underline{t}imation quality. 
The key idea of \FLAT is a novel unsupervised graphical model, called \FSPNend. It utilizes both independent and conditional factorization to adaptively model different levels of attributes correlations, and thus combines their advantages.  \FLAT supports efficient online probability computation in near linear time on the underlying \FSPN model, provides effective offline model construction and enables incremental model updates. It can estimate cardinality for both single table queries and multi-table join queries. Extensive experimental study demonstrates the superiority of \FLAT over existing \CE methods: \FLAT achieves 1–5 orders of magnitude better accuracy, 1–3 orders of magnitude faster probability computation speed and 1–2 orders of magnitude lower storage cost. We also integrate \FLAT into Postgres to perform an end-to-end test. It improves the query execution time by $12.9\%$ on the well-known IMDB benchmark workload, which is very close to the optimal result $14.2\%$ using the true cardinality.
\end{abstract}

\maketitle

\pagestyle{\vldbpagestyle}
\begingroup\small\noindent\raggedright\textbf{PVLDB Reference Format:}\\
\vldbauthors. \vldbtitle. PVLDB, \vldbvolume(\vldbissue): \vldbpages, \vldbyear.\\
\href{https://doi.org/\vldbdoi}{doi:\vldbdoi}
\endgroup

\begingroup
\renewcommand\thefootnote{}\footnote{\noindent
$\#$ The first two authors contribute equally to this paper. \\
$*$ Corresponding author. \\
\rule{\linewidth}{0.5pt}
This work is licensed under the Creative Commons BY-NC-ND 4.0 International License. Visit \url{https://creativecommons.org/licenses/by-nc-nd/4.0/} to view a copy of this license. For any use beyond those covered by this license, obtain permission by emailing \href{mailto:info@vldb.org}{info@vldb.org}. Copyright is held by the owner/author(s). Publication rights licensed to the VLDB Endowment. \\
\raggedright Proceedings of the VLDB Endowment, Vol. \vldbvolume, No. \vldbissue\ %
ISSN 2150-8097. \\
\href{https://doi.org/\vldbdoi}{doi:\vldbdoi} \\
}
\addtocounter{footnote}{-1}
\endgroup


\vspace{-1.5em}
\section{Introduction}
\label{sec: intro}

\emph{Cardinality estimation} (\textsf{CardEst}) is a key component of query optimizers in modern database management systems (DBMS) and analytic engines~\cite{armbrust2015spark,sethi2019presto}. Its purpose is to estimate the result size of a SQL query before its actual execution, thus playing a central role in generating high-quality query plans. 

Given a table $T$ and a query $Q$, estimating the cardinality of $Q$ is equivalent to computing $P$---the probability of records in $T$ satisfying $Q$. Therefore, the core task of \CE is to condense $T$ into a model $M$ to compute $P$. In general, such models could be obtained in two ways: \emph{query-driven} and \emph{data-driven}.
Query-driven approaches learn functions mapping a query $Q$ to its predicted probability $P$, so they require large amounts of executed queries as training samples. They only perform well if future queries follow the same distribution as the training workload. 
Data-driven approaches learn unsupervised models of $\Pr(T)$---the joint probability density function (PDF) of attributes in $T$. As they can generalize to unseen query workload, data-driven approaches receive more attention and are widely used for \CEend. 


\myskip
\noindent{\underline{\textbf{Challenge and Status of \CEend.}}} 
\revise{An effectual \CE method should satisfy three criteria~\cite{ioannidis1991propagation,tzoumas2011lightweight, Volcano1993The,wang2020ready}, namely high estimation accuracy, fast inference time and lightweight storage overhead, at the same time.} Existing methods have made some efforts in finding trade-offs between the them. However, they still suffer from one or more deficiencies when modeling real-world complex data.  

In a nutshell, there exist three major strategies for building unsupervised models of $\Pr(T)$ on data table $T$.
The first strategy directly compresses and stores all entries in $\Pr(T)$~\cite{gunopulos2005selectivity, poosala1997selectivity}, whose storage overhead is intractable and the lossy compression may significantly impact estimation accuracy.
The second strategy utilizes sampling~\cite{leis2017cardinality, zhao2018random} or kernel density based methods~\cite{kiefer2017estimating, heimel2015self}, where samples from $T$ are fetched on-the-fly to estimate probabilities. 
For high-dimensional data, they may be either inaccurate
without enough samples or inefficient due to a large sample size.

The third strategy, factorization based methods, 
is to decompose $\Pr(T)$ into multiple low-dimensional PDFs $\Pr(T')$ such that their suitable combination can approximate $\Pr(T)$. However, existing methods often fail to balance the three criteria. Some methods, including deep auto-regression~\cite{yang2020neurocard, yang2019deep,hasan2019multi} and Bayesian Network~\cite{tzoumas2011lightweight, getoor2001selectivity}, can losslessly decompose $\Pr(T)$ using \emph{conditional factorization}. However, their probability computation speed is reduced drastically. Other methods, such as \revise{1-D histogram}~\cite{selinger1979access} and sum-product network~\cite{hilprecht2019deepdb}, assume global or local \emph{independence} between attributes to decompose $\Pr(T)$. They attain high computation efficiency but their estimation accuracy is low when the independence assumption does not hold. We present a detailed analysis of existing data-driven \CE methods in Section~2.

\myskip
\noindent{\underline{\textbf{Our Contributions.}}}
In this paper, we address the \CE problem more comprehensively in order to satisfy all three criteria. We absorb the advantages of existing models and design a novel graphical model, called \textsf{\underline{f}actorize-\underline{s}um-\underline{s}plit-\underline{p}roduct} \underline{n}etwork (\FSPNend).
 Its key idea to \emph{adaptively} decompose $\Pr(T)$ according to the dependence level of attributes. 
Specifically, the joint PDF of highly and weakly correlated attributes will be losslessly separated by conditional factorization and modeled accordingly. 
The joint PDF of highly correlated attributes can be easily modeled as a multivariate PDF. For the weakly correlated attributes, their joint PDF is split into multiple small regions where attributes are mutually independent in each.
We prove that \FSPN subsumes \revise{1-D histogram}, sum-product network and Bayesian network, and leverages their advantages.

Based on the \FSPN model, we propose a \CE method called \FLATend, which is \emph{\underline{f}ast}, \emph{\underline{l}ightweight} and \emph{\underline{a}ccurate}. On a single table, \FLAT applies an effective offline method for the structure construction of \FSPN and an efficient online probability computation method using the \FSPNend. 
The probability computation complexity of \FLAT is almost linear w.r.t.~the number of nodes in \FSPNend. Moreover, \FLAT enables fast incremental updates of the \FSPN model.

For multi-table join queries, \FLAT uses a new framework, which is more general and applicable than existing work~\cite{hilprecht2019deepdb,yang2019deep,hasan2019multi,kipf2018learned}. In the offline phase, \FLAT clusters tables into several groups and builds an \FSPN for each group. In the online phase, \FLAT combines the probabilities of sub-queries in a fast way to get the final result.

In our evaluation, \FLAT achieves state-of-the-art performance on both single table and multi-table cases in comparison with all existing methods~\cite{kipf2018learned,yang2019naru,yang2020neurocard,hilprecht2019deepdb,tzoumas2011lightweight,poosala1997selectivity, kiefer2017estimating, leis2017cardinality}. On single table, \FLAT achieves up to $1$--$5$ orders of magnitude better accuracy, $1$--$3$ orders of magnitude faster probability computation speed (near $0.2ms$) and $1$--$2$ orders of magnitude lower storage cost (only tens of KB). On the \textit{JOB-light} benchmark~\cite{leis2018query, leis2015good} and a more complex crafted multi-table workload, \FLAT also attains the highest accuracy and an order of magnitude faster computation time (near $5ms$), while requiring only $3.3$MB storage space.
We also integrate \FLAT into Postgres. It improves the average end-to-end query time by $12.9\%$ on the benchmark workload, which is very close to the optimal result $14.2\%$ using the true cardinality.
This result confirms with a positive answer to the long-existing question whether and how much a more accurate \CE can improve the query plan quality~\cite{perron2019learned}. In addition, we have deployed \FLAT in the production environment of our company. We also plan to release to the community an open-source implementation of \FLATend.

In summary, our main contributions are listed as follows:

1) We analyze in detail the status of existing data-driven \CE methods in terms of the above three criteria (in Section~2). 

2) We present \FSPNend, a novel unsupervised graphical model, which combines the advantages of existing methods in an adaptive manner (in Section~3).

3) We propose \FLATend, a \CE method with fast probability computation, high estimation accuracy and low storage cost, on both single table and multi-table join queries (in Section~4 and 5).

4) We conduct extensive experiments and end-to-end test on Postgres to demonstrate the superiority and practicality of our proposed methods (in Section~6).


\section{$\!\!\!\!$ Problem Definition and Background}
\label{sec: prebkg}

In this section, we formally define the \CE problem and analyze the status of data-driven \CE
methods. Based on the analysis, we summarize some key findings that inspire our work.

\myskip
\noindent{\underline{\textbf{\CE Problem.}}}
Let $T$ be a table with a set of $k$ attributes $A = \{A_1, A_2, \dots, A_k \}$.
\revise{$T$ could either be a single or a joined table.} Each attribute $A_i$ in $T$ is assumed to be either categorical, so that values can be mapped to integers, or continuous.
Without loss of generality, we assume that the domain of $A_i$ 
is $[LB_i, UB_i]$. 

In this paper, we do not consider ``LIKE'' queries on strings.
Any selection query $Q$ on $T$ may be represented in canonical form: $Q = (A_1 \in [L_1, U_1] \wedge A_2 \in [L_2, U_2] \wedge \cdots  \wedge A_k \in [L_k, U_k])$, where $LB_i \leq L_i \leq U_i \leq UB_i$ for all $i$.
W.l.o.g., the endpoints of each interval can also be open. We call $Q$ a \emph{point query} if $L_i = U_i$ for all $i$ and \emph{range query} otherwise. 
If $Q$ has no constraint on the left or right hand side of $A_i$, we simply set $L_i = LB_i$ or $U_i = UB_i$, respectively. For any query $Q'$ where the constraint of an attribute $A_i$ contains several intervals, we may split $Q'$ into multiple queries satisfying the above form.

Let $\Card(T, Q)$ denote the exact number of records in $T$ satisfying all predicates in $Q$. Generally, the \CE problem asks to estimate the value of $\Card(T, Q)$ as accurately as possible without executing $Q$ on $T$.
 \CE is often modeled and solved from a statistical perspective. 
 We can regard each attribute $A_i$ in $T$ as a random variable.
 The table $T$ essentially represents a set of i.i.d.~records sampled from the joint PDF $\Pr_{\T}(A) = \Pr_{\T}(A_1, A_2, \dots, A_k)$. For any query $Q$, let $\Pr_{\T}(Q)$ denote the probability of records in $T$ satisfying $Q$. We have 
$\Card(T, Q) = \Pr\nolimits_{\T}(Q) \cdot |T|$.
Therefore, estimating $\Card(T, Q)$ is equivalent to estimating the probability $\Pr_{\T}(Q)$. Unsupervised \CE solves this problem in a purely data-driven fashion, which can be formally stated as follows: 

\noindent{\textbf{Offline Training:}} Given a table $T$ with a set $A$ of attributes as input, output a model  $\widehat{\Pr}_{\T}(A)$ for $\Pr_{\T}(A)$ such that $\widehat{\Pr}_{\T}(A) \approx \Pr_{\T}(A)$.

\noindent{\textbf{Online Probability Computation:}} Given the model $\widehat{\Pr}_{\T}(A)$ and a query $Q$ as input, output $\widehat{\Pr}_{\T}(Q) \cdot |T|$ as the estimated cardinality. 

\myskip
\noindent{\underline{\textbf{Data-Driven \CE Methods Analysis.}}}
We use three criteria, namely \emph{model accuracy}, \emph{probability computation speed} and \emph{storage overhead}, to analyze existing methods. The results are as follows:

1) \textsf{\underline{Lossy FullStore}}~\cite{gunopulos2005selectivity} stores all entries in $\Pr_{\T}(A)$ using compression techniques, whose storage grows exponentially in the number of attributes and becomes intractable~\cite{yang2019deep,yang2020neurocard}.

2) \textsf{\underline{Sample and Kernel-based methods}}~\cite{leis2017cardinality, zhao2018random,kiefer2017estimating, heimel2015self} do not store $\Pr_{\T}(A)$ but rather sample records from $T$ on-the-fly, or use average kernels centered around sampled points to estimate $\Pr_{\T}(Q)$. For high-dimensional data, they may be either inaccurate without enough samples, or inefficient due to a large sample size.

\vspace{0.5em}
Alternatively, a more promising way is to \emph{factorize} $\Pr_{\T}(A)$ into multiple low-dimensional PDFs $\Pr_{\T}(A')$ such that: 1) $|A'| << |A|$ so $\Pr_{\T}(A')$ is easier to store and model; and 2) a suitable combination, e.g. multiplication, weighted sum and etc, of $\Pr_{\T}(A')$ approximates $\Pr_{\T}(A)$. Some representative methods are listed in the following:

\indent 3) \textsf{\underline{\revise{1-D Histogram}}}~\cite{selinger1979access} assumes all attributes are mutually independent, so that $\widehat{\Pr}_{\T}(A) = \prod_{i = 1}^{k} \widehat{\Pr}_{\T}(A_i)$. Each $\widehat{\Pr}_{\T}(A_i)$ is built as a (cumulative) histogram, so $\widehat{\Pr}_{\T}(Q)$ may be obtained in $O(|A|)$ time. However, the estimation errors may be high, since correlations between attributes are ignored. 

\indent 4) \textsf{\underline{\revise{M-D Histogram}}}
\revise{~\cite{poosala1997selectivity, deshpande2001independence, gunopulos2000approximating, wang2003multi} builds multi-dimensional histograms to model the dependency of attributes. They identify subsets of correlated attributes using models such as Markov network, build histograms on each subset and assume the independence across different subsets. It improves the accuracy but the decomposition is still lossy. Meanwhile, it is space consuming.
}
\\
\indent 5) \textsf{\underline{Deep Auto-Regression (DAR)}}~\cite{yang2020neurocard, yang2019deep,hasan2019multi} decomposes the joint PDF according to the chain rule, i.e., $\Pr_{\T}(A) \! = \! \Pr_{\T}(A_1) \cdot \prod_{i=2}^{k} \! \Pr_{\T}(A_i| \break A_1, A_2,  \dots, A_{i - 1})$. Each conditional PDF can be parametrically modeled by a deep neural network (DNN). While the expressiveness of DNNs allows $\Pr_{\T}(A)$ to be approximated well, probability computation time and space cost increase with the width and depth of the DNN. Moreover, for range query $Q$, computing $\Pr_{\T}(Q)$ requires averaging the probabilities of lots of sample points in the range. Thus, the probability computation on \textsf{DAR} is relatively slow. \\
\indent 6) \textsf{\underline{Bayesian Network (BN)}} ~\cite{tzoumas2011lightweight, getoor2001selectivity, chow1968approximating} models the dependence structure between all attributes as a directed acyclic graph and assumes that each attribute is conditionally independent of the remaining attributes given its parents. The probability $\Pr_{\T}(A)$ is factorized as $\Pr_{\T}(A) = \prod_{i = 1}^{k} \Pr_{\T}(A_i | A_{\text{pa}(i)})$, where $\text{pa}(i)$ is the parent attributes of $A_i$ in \textsf{BN}.  Learning the \textsf{BN} structure from data and probability computation on \textsf{BN} are both NP-hard~\cite{scanagatta2019survey, dagum1993approximating}.\\
\indent 7) \textsf{\underline{Sum-Product Network (SPN)}}~\cite{hilprecht2019deepdb}
approximates $\Pr_{\T}(A)$ using several local and simple PDFs. An \textsf{SPN} is tree structure where each node stands for an estimated PDF $\widehat{\Pr}_{\T'}(A')$ of the attribute subset $A'$ on record subset $T' \subseteq T$~\cite{poon2011sum}. The root node represents $\widehat{\Pr}_{\T}(A)$. 
Each inner node is: 1) a \textsf{sum} node which splits all records (rows) in $T'$ into $T_{i}'$ on each child such that $\widehat{\Pr}_{\T'}(A') = \sum_{i} w_i \widehat{\Pr}_{\T_i'}(A')$ with weights $w_i$;
or 2) a \textsf{product} node which splits attributes (columns)  in $A'$ on each child as $\widehat{\Pr}_{\T'}(A') = \prod_{j} \widehat{\Pr}_{\T'}(A_j')$ when all $A_j'$ are mutually independent in $T'$. Each leaf node then maintains a (cumulative) PDF on a singleton attribute. The probability $\widehat{\Pr}_{\T}(Q)$ can be computed in a bottom-up manner using the \textsf{SPN} node operations for both point and range queries. The storage overhead and probability computation cost are linear in the number of nodes of \textsf{SPN}. 

The performance of \textsf{SPN} heavily relies on the
local independence assumption. When it holds, the generated \textsf{SPN} is compact and exhibits superiority over other methods~\cite{hilprecht2019deepdb,yang2020neurocard}. 
However, real-world data often possesses substantial skew and strong correlations between attributes~\cite{tzoumas2011lightweight}. In this situation, \textsf{SPN} can not split these attributes using the \textsf{product} operation and might repeatedly apply the \textsf{sum} operation to split records into extremely small volumes~\cite{SPN_expressive}, i.e., $|T'| = 1$. This would heavily increase the \textsf{SPN} size, degrade its efficiency and make the model inaccurate~\cite{desana2020sum, SPN_expressive}.

\myskip
\noindent{\underline{\textbf{Inspirations.}}}
\revise{Based on the analysis, there does not exist a comprehensively effectual \CE method since each method only utilizes one factorization approach.} However, independent factorization has low storage cost and supports fast inference but may incur huge estimation errors; conditional factorization can accurately decompose the PDF but the inference is costly. This leads to our key question: \emph{if we could, in an adaptive manner, apply both kinds of factorization, would it be possible to obtain a \CE method that can simultaneously satisfy all three criteria}?
We answer this question affirmatively with a new unsupervised model, called \textsf{factorize-split-sum-product network (FSPN)}, which integrates the strength of both factorization approaches.



\vspace{-0.5em} 

\section{The \textsf{FSPN} Model}
\label{sec: fspn}

In this section, we present \FSPNend, a new tree-structured graphical model representing the joint PDF of a set of attributes in an \emph{adaptive} manner. We first explain the key ideas of \FSPN with an example and then present its formal definition. Finally, we compare \FSPN with aforementioned models.

\begin{figure*}
	\centering
	\includegraphics[width=0.88
	\linewidth]{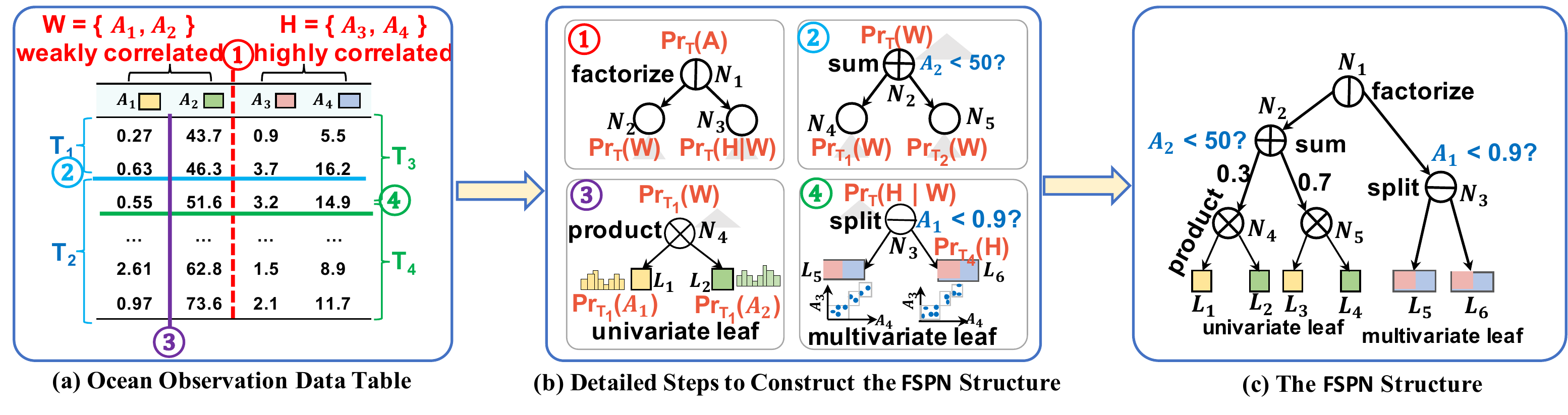}
	\vspace{-1em}
	\caption{An ocean observation data table and its corresponding \FSPNend.}
	\label{fig: fspnexample}
	\vspace{-1em}
\end{figure*}

\myskip
\noindent{\underline{\textbf{Key Ideas of \FSPNend.}}}
\FSPN can factorize attributes with different dependence levels accordingly. 
The conditional factorization approach is used to split highly and weakly correlated attributes.
Then, highly correlated attributes are directly modeled together while weakly correlated attributes are recursively approximated using the 
independent factorization approach. Figure~\ref{fig: fspnexample}(a) gives an example of table $T$ with a set $A$ of four attributes \textsl{water turbidity} ($A_1$), \textsl{temperature} ($A_2$), \textsl{wave height} ($A_3$) and \textsl{wind force} ($A_4$). We elaborate the process to construct its \FSPN in Figure~\ref{fig: fspnexample}(b) as follows:

At first, we examine the correlations between each pair of attributes in $T$. $A_3$ and $A_4$ are globally highly correlated, so they can not be decomposed as independent attributes unless we split $T$ into extremely small clusters as \textsf{SPN}. Instead, we can losslessly separate them  from other attributes as early as possible and process each part respectively.
Let $H = \{A_3, A_4\}$ and $W = \{A_1, A_2\}$. We apply the conditional factorization approach and factorize $\Pr_{\T}(A) = \Pr_{\T}(W) \cdot \Pr_{\T}(H | W)$ (as node $N_1$ in step~\textcircled{1}). $\Pr_{\T}(W)$ and $\Pr_{\T}(H | W)$ are then modeled in different ways. 

The two attributes $A_1$ and $A_2$ in $W$ are not independent on $T$ but they are weakly correlated. 
Thus, we can utilize the independent factorization approach on small subsets of $T$. 
In our example, if we split all records in $T$ into $T_1$ and $T_2$ based on whether $A_1$ is less than $50$ (as node $N_2$ in step~\textcircled{2}), $A_1$ and $A_2$ are independent on both $T_1$ and $T_2$. This situation is called \emph{contextually independent}, where $T_1$ and $T_2$ refer to the specific context. Since $\Pr_{\T_1}(W) = \Pr_{\T_1}(A_1) \cdot \Pr_{\T_1}(A_2)$ (as node $N_4$ in step~\textcircled{3}), we then simply use two univariate PDFs (such as histograms in leaf nodes $L_1$ and $L_2$ in step~\textcircled{3}) to model $\Pr_{\T_1}(A_1)$ and $\Pr_{\T_1}(A_2)$ on $T_1$, respectively. Similarly, we also model $\Pr_{\T_2}(W) = \Pr_{\T_2}(A_1) \cdot \Pr_{\T_2}(A_2)$ on $T_2$ (as node $N_5$).

For the conditional PDF $\Pr_{\T}(H | W)$, we do not need to specify $\Pr(H | w)$ for each value $w$ of $W$. Instead, we can recursively split $T$ into multiple regions $T_j$ in terms of $W$ such that  $H$ is  independent of $W$ in each context $T_j$, i.e., $\Pr_{\T_j}(H) = \Pr_{\T_j}(H | W)$. At this time, for any value $w$ of $W$ falling in the same region, $\Pr_{\T_j}(H | w)$ stays the same, so we only need to maintain $\Pr_{\T_j}(H)$ for each region. We refer to this situation as \emph{contextual condition removal}. 
In our example, we split $T$ into $T_3$ and $T_4$ (as nodes $N_3$ in step~\textcircled{4}) by whether the condition attribute $A_1$ is less than $0.9$. $W$ is independent of $H$ on each leaf node region, so we only need to model $\Pr_{\T_3}(H)$ and $\Pr_{\T_4}(H)$. Thus, we model them as
 two multivariate leaf nodes $L_5$ and $L_6$ in step~\textcircled{4}. Note that, attribute values in $H$ are interdependent and their joint PDFs $\Pr_{\T_3}(H)$ and $\Pr_{\T_4}(H)$ are sparse in the two-dimensional space, so they are easy modeled as a multivariate PDF. 

Finally, we obtain an \FSPN in Figure~\ref{fig: fspnexample}(c) containing 11 nodes, where 5 inner nodes represent different operations to split data and 6 leaf nodes keep PDFs for different parts of the original data.

\myskip
\noindent{\underline{\textbf{Formulation of \FSPNend.}}} 
Let $\F$ denote a \FSPN modeling the joint PDF $\Pr_{\T}(A)$ for records $T$ with attributes $A$. $\F$ is a tree structure. Each node $N$ in $\F$ is a 4-tuple $(A_{\N}, C_{\N}, T_{\N}, O_{\N})$ where:

\indent $\bullet$ 
$T_{\N} \subseteq T$ represents a set of records where the PDF is built on. It is called the \emph{context} of node $N$. 

\indent $\bullet$ 
$A_{\N}, C_{\N} \subseteq A$ represent two set of attributes. 
We call $A_\N$ and $C_\N$ the \emph{scope} and \emph{condition} of node $N$, respectively. If $C_\N = \emptyset$, $N$ represents the PDF $\Pr_{\T_{\N}}(A_\N)$;
otherwise, it represents the conditional PDF $\Pr_{\T_{\N}}(A_{\N} | C_{\N})$.
The root of $\F$, such as $N_1$ in Figure~\ref{fig: fspnexample}(c), is a node with $A_\N = A$, $C_\N = \emptyset$ and $T_\N = T$ representing the joint PDF $\Pr_{\T}(A)$.

\indent $\bullet$ 
$O_{\N}$ stands for the \emph{operation} specifying how to split data to generate its children in different ways:\\
\indent 1) A \underline{\textsf{Factorize} (\textcircled{$|$})}
node, such as $N_1$ in step~\textcircled{1}, splits highly correlated attributes from the remaining ones by conditional factorization only when $C_\N = \emptyset$.
Let $H \subseteq A_\N$ be a subset of highly correlated attributes. It generates the left child  $N_{\L} = (A_\N - H, \emptyset, T_\N, O_{\L})$ and the right child $N_{\R} =  (H, A_\N - H, T_\N, O_{\R})$. We have $\Pr_{\T_{\N}}(A_\N) = \Pr_{\T_{\N}}(A_\N - H) \cdot \Pr_{\T_{\N}}(H | A_\N - H)$. \\
\indent 2) A \underline{\textsf{Sum} (\textcircled{$+$})} node, such as $N_2$ in step~\textcircled{2}, splits the records in $T_\N$ in order to enforce contextual independence only when $C_\N = \emptyset$. We partition $T_\N$ into subsets $T_1, T_2, \dots, T_n$.
For each $1 \leq i \leq n$, $N$ generates the child $N_{i} = (A_\N, \emptyset, T_i, O_i)$ with weight $w_i = |T_i|/|T_\N|$. We can regard $N$ as a mixture of models on all of its children, i.e., $\Pr_{\T_{\N}}(A_\N ) = \sum_{i = 1}^{n} w_i \Pr_{\T_i}(A_\N)$, where $w_i$ represents the proportion of the $i$-th subset. \\
\indent 3) A \underline{\textsf{Product} (\textcircled{$\times$})} node, such as $N_4$ in step~\textcircled{3}, splits the scope $A_\N$ of $N$ only when $C_\N = \emptyset$ and contextual independence holds. Let $A_1, A_2, \dots, A_m$ be the mutually independent partitions of $A_\N$. $N$ generates children $N_j = (A_j, \emptyset, T_\N, O_j)$ for all $1 \leq j \leq m$ such that $\Pr_{\T_{\N}}(A_\N) = \prod_{j = 1}^{m} \Pr_{\T_{\N}}(A_j)$. \\
\indent 4) A \underline{\textsf{Split} (\textcircled{$-$})} node, such as $N_3$ in step~\textcircled{4}, partitions the records $T_\N$ into disjoint subsets $T_1, T_2, \dots, T_d$ only when $C_\N \neq \emptyset$. For each $1 \leq i \leq d$, $N$ generates the child $N_{i} = (A_\N, C_\N, T_i, O_i)$. Note that for any value $c$ of $C_\N$, there exists exactly one $j$ such that $c$ falls in the region of  $T_j$. 
The semantic of \textsf{split} is different from \textsf{sum}. 
The \textsf{split} node divides a large model of $\Pr_{T_\N}(A_\N | C_\N)$ into several parts by the values of $C_\N$. Whereas, the \textsf{sum} node decomposes a large model of $\Pr_{T_\N}(A_\N)$ to small models on the space of $A_\N$. 
\\
\indent 5) A \underline{\textsf{Uni-leaf} ($\square$)} node, such as  $L_1$ and $L_2$ in step~\textcircled{3}, keeps the univariate PDF $\Pr_{T_\N}(A_\N)$, such as histogram or Gaussian mixture model, only when $|A_\N| = 1$ and $C_\N = \emptyset$. \\
\indent 6) A \underline{\textsf{Multi-leaf} ($\square \! \square$)} node, such as $L_5$ and $L_6$ in step~\textcircled{4}, maintains the multivariate PDF $\Pr_{T_\N}(A_\N)$ only when $C_\N \neq \emptyset$ and $A_\N$ is independent of $C_\N$ on $T_\N$.

The above operations are recursively used to construct $\F$ with three constraints: 
1) for a \textsf{factorize} node, the right child must be a \textsf{split} node or \textsf{multi-leaf}; the left child can be any type in \textsf{sum}, \textsf{product}, \textsf{factorize} and \textsf{uni-leaf};
2) the children of a \textsf{sum} or \textsf{product} node could be any type in \textsf{sum}, \textsf{product}, \textsf{factorize} and \textsf{uni-leaf};
and 3) the children of a \textsf{split} node can only be \textsf{split} or \textsf{multi-leaf} nodes.

\myskip
\noindent{\underline{\textbf{Differences with \textsf{SPN}.}}}
As the name suggests, \FSPN is inspired by SPN and its successful application in \CE~\cite{hilprecht2019deepdb}. However, \FSPN differs from \textsf{SPN} in two fundamental aspects.
First, in terms of the underlying key ideas, \FSPN tries to adaptively model attributes with different levels of dependency, which is not considered in \textsf{SPN}. 
Second, in terms of the fundamental design choices,
\FSPN can split weakly and highly correlated attributes, and models each class differently: 1) weakly correlated attributes are modeled by \textsf{sum} and \textsf{product} operations; and 2) for highly correlated attributes, \FSPN uses \textsf{split} and \textsf{multi-leaf} nodes. \textsf{SPN} only uses the first technique on all attributes. As per our analysis in Section~2, this can generate a large structure since local independence can not easily hold. 

Moreover, a simple extension of \textsf{SPN} with \textsf{multi-leaf} nodes also seems unlikely to mitigate its inherent limitations. This is because multi-leaf nodes can only efficiently model highly correlated attributes, as their joint PDF can be easily reduced to and modeled in a low dimensional space. Otherwise, their storage cost grows exponentially so the model size would be very large. \FSPN guarantees that multi-leaf nodes are only applied on highly correlated attributes, 
whereas \textsf{SPN} and its extensions lack such mechanism. 
\revise{Our experimental results in Section~6.1 exhibit that the model size of \textsf{SPN} with \textsf{multi-leaf} nodes are much larger than \FSPN and may exceed the memory limit on highly correlated table.}

\myskip
\noindent{\underline{\textbf{Generality of \FSPNend.}}}
We show that FSPN \emph{generalizes} \revise{\textsf{1-D Histogram}}, \textsf{SPN} and \textsf{BN} models. 
First, when all attributes are mutually independent, 
\FSPN becomes \revise{1-D \textsf{Histogram}}.
Second, \FSPN degenerates to \textsf{SPN} by disabling the \textsf{factorize} operation.
Third, \FSPN could equally represent a \textsf{BN} model on discrete attributes by iteratively factorizing each attribute having no parents from others. We put the transformation process in Appendix~A.1. Based on it, we obtain Lemma~1 (proved in Appendix~A.2) stating that the \FSPN is no worse than \textsf{SPN} and \textsf{BN} in terms of expressive efficiency.

\textbf{\textit{Lemma~1}}
\textit{Given a table $T$ with attributes $A$, if the joint PDF $\Pr_{\T}(A)$ is represented by an \textsf{SPN} $\mathcal{S}$ or a \textsf{BN} $\mathcal{B}$ with space cost $O(M)$, then there exists an \FSPN $\mathcal{F}$ that can equivalently model $\Pr_{\T}(A)$ with no more than $O(M)$ space.}


\section{Single Table C\lowercase{ard}E\lowercase{st}  Method}
\label{sec: single}

In this section, we propose \FLATend, a \underline{f}ast, \underline{l}ightweight and \underline{a}ccurate \textsf{CardEs\underline{t}} algorithm built on \FSPNend. We first introduce how \FLAT computes the probability on \FSPN online in Section~4.1. Then, we show how \FLAT constructs the \FSPN from data offline in Section~4.2.
Finally, we discuss how \FLAT updates the model in Section~4.3.

\subsection{Online Probability Computation}
\label{sec: single-1}

\FLAT can obtain the probability (cardinality) of any query $Q$ in a recursive manner on \FSPNend. We first show the basic strategy of probability computation with an example, and then present the detailed algorithm and analyze its complexity.

\begin{figure}[!t]
	\resizebox{0.75\linewidth}{!}{
		\begin{tabular}{c|cccc}
			\hline
			\rowcolor{mygrey}
			\bf Range & \boldmath{$A_1$} {\color{mywt} $\blacksquare$} & \boldmath{$A_2$} {\color{mytp} $\blacksquare$} & \boldmath{$A_3$} {\color{mywh} $\blacksquare$}  & \boldmath{$A_4$} {\color{mywf} $\blacksquare$} \\ \hline
			Bound & [0, 10] & [0, 100] & [0, 100] & [0, 100] \\ \hline
			Leaf $L_5$ & [0, 0.9) & [0, 100] & [0, 100] & [0, 100]\\
			Leaf $L_6$ & [0.9, 10] & [0, 100] & [0, 100] & [0, 100 \\ \hline \hline
			\rowcolor{myblue}
			\textbf{Query} \boldmath$Q$ & [0.6, 1.4] & [35, 65] & [2, 3] & [60, 70]\\ 
			Query $Q_1$ & [0.6, 0.9) & [35, 65] & [2, 3] & [60, 70] \\ 
			Query $Q_2$ & [0.9, 1.4] & [35, 65] & [2, 3] & [60, 70] \\ 
			\hline
		\end{tabular} 
	}
	\includegraphics[width = 0.75\linewidth]{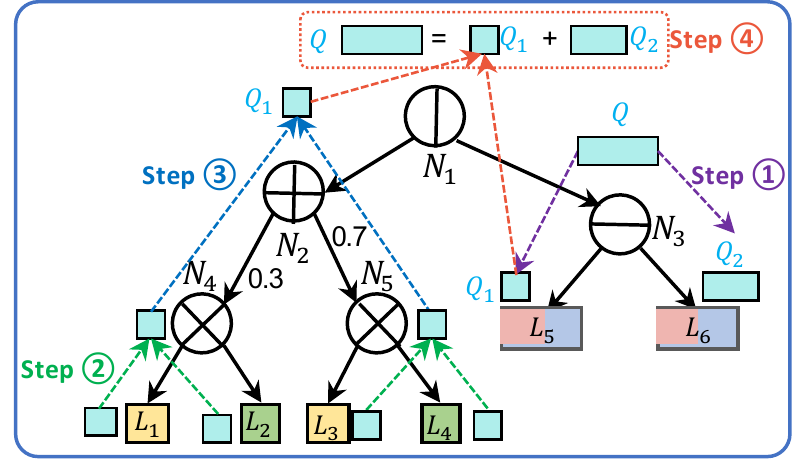}
	\vspace{-1em}
	\caption{An example of the \FLAT probability computation.}
	\label{fig: fspninferexp}
	\vspace{-1em}
\end{figure}

\myskip
\noindent{\underline{\textbf{Basic Strategy.}}}
As stated in Section~\ref{sec: prebkg}, the query $Q$ can be represented in canonical form: $Q = (A_1 \in [L_1, U_1] \wedge A_2 \in [L_2, U_2] \wedge \cdots  \wedge \break A_k \in [L_k, U_k])$, where $L_i \leq A_i \leq U_i$ is the constraint on attribute $A_i$.
Obviously, $Q$ represents a \emph{hyper-rectangle} range in the attribute space whose probability needs to be computed. In Figure~\ref{fig: fspninferexp}, we give an example query $Q$ on the \FSPN in Figure~\ref{fig: fspnexample}(c).

First, considering the root node $N_1$, computing the probability of $Q$ on this  
\textsf{factorize} node is a non-trivial task. For each point $r \in Q$, we can obtain its probability $\Pr_{r}(A_1, A_2)$ from node $N_2$ and the conditional probability $\Pr_{r}(A_3, A_4|A_1, A_2)$ from node $N_3$. However, for different $r$, $\Pr_{r}(A_3, A_4|A_1, A_2)$ is modeled by different PDFs on \textsf{multi-leaf} nodes $L_5$ or $L_6$ of $N_3$. Thus,  we must split $Q$ into two regions to compute the probability of $Q$ (as step~\textcircled{1} in Figure~\ref{fig: fspninferexp}). To this end, we push $Q$ onto $N_3$, whose splitting rule on the condition attributes ($A_1 < 0.9$) would divide $Q$ into two hyper-rectangle ranges $Q_1$ and $Q_2$ on \textsf{multi-leaf} nodes $L_5$ or $L_6$, respectively. For $Q_1$ (or $Q_2$), the probability $\Pr(A_3, A_4|A_1, A_2) = \Pr(A_3, A_4)$ can be directly obtained from the multivariate PDF on $L_5$ (or $L_6$).

Then, we can compute the probability $\Pr(A_1, A_2)$ for each region $Q_1$ and $Q_2$ from $N_2$. Obviously, for the \textsf{sum} node (e.g.~$N_2$) and \textsf{product} node (e.g.~$N_4$), the probability of each region can be recursively obtained by summing (as step~\textcircled{3}) or multiplying (as step~\textcircled{2}) the probability values of its children, respectively. In the base case, the probability on the singleton attribute $A_1$ (or $A_2$) is obtained from the \textsf{uni-leaf} nodes $L_1$ and $L_3$ (or $L_2$ and $L_4$). Finally, since $\Pr(A_1, A_2)$ and $\Pr(A_3, A_4)$ are independent in $Q_1$ and $Q_2$, we can multiply and sum them together ((as step~\textcircled{4})) to obtain the probability of $Q$.

\myskip
\noindent{\underline{\textbf{Algorithm Description.}}}
Next, we describe the online probability computation algorithm \textsf{FLAT-Online}. It takes as inputs a \FSPN $\F$ modeling $\Pr_{\T}(A)$ and the query $Q$, and outputs $\Pr_{\T}(Q)$ on $\F$. Let $N$ be the root node of $\F$ (line 1). For any node $N'$ in $\F$, let $\F_{\N'}$ denote the \FSPN rooted at $N'$. \textsf{FLAT-Online} recursively computes the probability of $Q$ by the following rules:

\indent \underline{\textit{Rule~1 (lines~2--3)}}: 
Basically, if $N$ is a \textsf{uni-leaf} node, we directly return the probability of $Q$ on the univariate PDF of the attribute.

\indent \underline{\textit{Rule~2 (lines~4--11)}}: 
if $N$ is a \textsf{sum} node (lines~4--7) or a \textsf{product} node (lines~8--11), let $N_1, N_2, \dots, N_t$ be all of its children. We can further call \textsf{FLAT-Online} on $\F_{\N_i}$ for each $1 \leq i \leq t$ to obtain the probability on the
PDF represented by each child. Then, node $N$ 
computes a weighted sum (for \textsf{sum} node) or
multiplication (for \textsf{product} node) of these probabilities.

\indent \underline{\textit{Rule~3 (lines~12--18)}}: 
if $N$ is a \textsf{factorize} node, let $LC$ and $RC$ be its left and right child modeling $\Pr_{\T}(W)$ and $\Pr_{\T}(H|W)$, respectively. All descendants of $RC$ are \textsf{split} or \textsf{multi-leaf} nodes. Let $L_1, L_2, \dots, L_t$ be all \textsf{multi-leaf} descendants of $RC$. We assume that each \textsf{split} node divides the attribute domain space in a grid manner, which is ensured by the \FSPN structure construction method in Section~4.2.
Then, each $L_i$ maintains a multivariate PDF on a hyper-rectangle range specified by all \textsf{split} nodes on the path from $RC$ to $L_i$. 
Based on these ranges, we can divide the range of query $Q$ into $Q_1, Q_2, \dots, Q_t$. For each $Q_i$, the probability $h_i$ on highly correlated attributes $H$ could be directly obtained from $L_i$. The probability $w_i$ on attributes $W$ could be recursively obtained by calling \textsf{FLAT-Online} on $\F_{\L\C}$, the \FSPN rooted at $LC$, and $Q_i$. After that, since $H$ is independent of $W$ on the range of each $Q_i$, we sum all products $h_i w_i$ together as the probability of $Q$.

\begin{figure}[t]
	\scriptsize
	\rule{\linewidth}{1pt}
	\leftline{~~~~{\textbf{Algorithm} \textsf{FLAT-Online$(\F, Q)$}}}
	\vspace{-1em}
	\begin{algorithmic}[1]
		\STATE let $N$ be the root node of $\mathcal{F}$
		\IF{$N$ is \textsf{uni-leaf} node}
		\RETURN $\Pr_{\T}(Q)$ by the univariate PDF on the attribute modeled by $N$
		\ELSIF{$N$ is a \textsf{sum} node}
		\STATE let $N_1, N_2, \dots, N_t$ be the children of $N$ with weights $w_1, w_2, \dots, w_t$
		\STATE $p_i \gets \textsf{FLAT-Online}(\mathcal{F}_{\N_i}, Q)$ for each $1 \leq i \leq t$
		\RETURN $\sum_{i = 1}^{t} w_i p_i$
		\ELSIF{$N$ is a \textsf{product} node}
		\STATE let $N_1, N_2, \dots, N_t$ be the children of $N$
		\STATE $p_i \gets \textsf{FLAT-Online}(\mathcal{F}_{\N_i}, Q)$ for each $1 \leq i \leq t$
		\RETURN $\prod_{i = 1}^{t} p_i$
		\ELSE
		\STATE let $LC$ be the left child modeling $\Pr_{\T}(W)$ and $RC$ be the right child modeling $\Pr_{\T}(H | W)$
		\STATE let $L_1, L_2, \dots, L_t$ be all the  \textsf{multi-leaf} descendants of $RC$ 
		\STATE split $Q$ into $Q_1, Q_2, \dots, Q_t$ by ranges of $L_1, L_2, \dots, L_t$
		\STATE get $h_i$ of $Q_i$ on variables $H$ from the multivariate PDF on $L_i$ for each $1 \leq i \leq t$
		\STATE $w_i \gets \textsf{FLAT-Online}(\mathcal{F}_{\L\C}, Q_i)$ for each $1 \leq i \leq t$
		\RETURN $\sum_{i = 1}^{t} h_i w_i$
		\ENDIF
	\end{algorithmic}
	\vspace{-0.5em}
		\rule{\linewidth}{1pt}
		\vspace{-4em}
\end{figure}

\myskip
\noindent{\underline{\textbf{Complexity Analysis.}}}
We assume that, on each leaf node, the probability of any range can be computed in $O(1)$ time, which can be easily implemented by a cumulative histogram or Gaussian mixture functions. Let $n$ be the number of nodes in  \FSPNend. Let $f$ and $m$ be the number of \textsf{factorize} and \textsf{multi-leaf} nodes in \FSPNend, respectively. The maximum number of ranges to be computed on each node is $O(m^f)$, so the time cost of \textsf{FLAT-Online} is $O(m^f n)$.

\revise{By our empirical testing, the actual time cost of \textsf{FLAT-Online} is almost linear w.r.t.~the number of nodes in \FSPN for two reasons.}
First, \FSPN is compact on real-world data so both $f$ and $n$ are small. Second, the computation on many ranges in each node could be easily done in parallel. 
 In our testing, the speed of \textsf{FLAT-Online} is even near the histogram method and $1$--$3$ orders of magnitude faster than other methods (See Section~6.1).

\subsection{Offline Structure Construction}
\label{sec: single-2}

\begin{figure}[t]
	\includegraphics[width = 0.8\columnwidth]{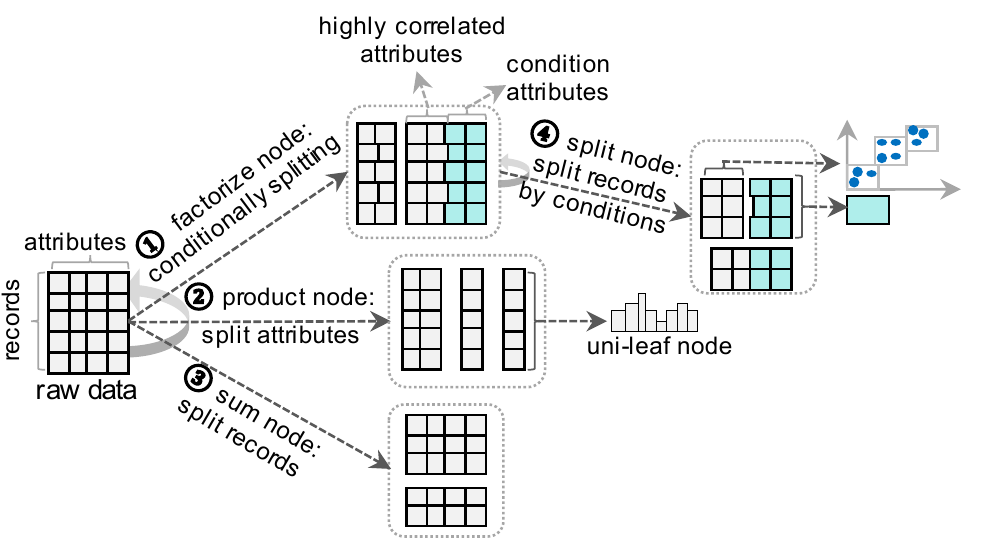}
	\vspace{-1.5em}
	\caption{\FLAT Structure Construction Process.}
	\label{fig: fspnstruc}
	\vspace{-2em}
\end{figure}

We present the detailed procedures to build an \FSPN in the algorithm \textsf{FLAT-Offline}. Its general process is shown in Figure~\ref{fig: fspnstruc}.
\textsf{FLAT-Offline} works in a top-down manner. Each node $N$ takes the scope attributes $A_\N$, the condition attributes $C_\N$ and the context of records $T_\N$ as inputs, and recursively decompose the joint PDF to build the \FSPN rooted at $N$. To build the \FSPN $\F$ modeling table $T$ with attributes $A$, we can directly call $\textsf{FLAT-Offline}(A, \emptyset, T)$. We briefly scan its main procedures as follows:

\underline{\textit{1. Separating highly correlated attributes with others (lines~2--8)}}: \break
when $C_\N = \emptyset$, \textsf{FLAT-Offline} firstly detects if
there exists a set $H$ of highly correlated attributes since the principle of \FSPN is to separate them with others as early as possible (step~\textcircled{1} in Figure~\ref{fig: fspnstruc}). We find $H$ by examining pairwise correlations, e.g.~RDC~\cite{lopez2013randomized}, between attributes and iteratively group attributes whose correlation value is larger than a threshold $\tau_{h}$. If $H \neq \emptyset$, we set $N$ to be a \textsf{factorize} node. The left child and right child of $N$ recursively call \textsf{FLAT-Offline} to model $\Pr_{\T_\N}(A_\N - H)$ and $\Pr_{T_\N}(H | A_\N - H)$, respectively.

\underline{\textit{2. Modeling weakly correlated attributes (lines~9--19)}}: 
if $C_\N = \emptyset$ and $H = \emptyset$, we try to split $\Pr_{\T_\N}(A_\N)$ into small regions such that attributes in $A_\N$ are locally independent. Specifically, if $|A_\N| = 1$, $N$ is a \textsf{uni-leaf} node (line~10). We call the \textsf{Leaf-PDF} procedure to model univariate PDF $\Pr_{\T_{\N}}(A_\N)$ (line~11) using off-the-shelf tools. In our implementation, we choose histograms~\cite{poosala1997selectivity} and parametric Gaussian mixture functions~\cite{rasmussen2000infinite} to model categorical and continuous attributes, respectively. 

Otherwise, we try to partition $A_\N$ into mutually independent subsets based on their pairwise correlations (step~\textcircled{2} in Figure~\ref{fig: fspnstruc}). Two attributes are regarded as independent if their correlation value is no larger than than a threshold $\tau_{l}$. If $A_\N$ can be split to 
mutually independent subsets $A_1, A_2, \dots, A_m$, we set $N$ to be a \textsf{product} node and call \textsf{FLAT-Offline} to model $\Pr_{T_\N}(A_i)$ for each $1 \leq i \leq m$ (lines~12--14). If not, the local independency does not exist, so we need to split the data (step~\textcircled{3} in Figure~\ref{fig: fspnstruc}). 
\revise{Similar to~\cite{gens2013learning}, we 
apply a clustering method, such as $k$-means~\cite{krishna1999genetic}, to cluster $T_\N$ to $T_1, T_2, \dots, T_n$ according to $A_\N$ (line~17). The records in the same cluster are similar, so the corresponding PDF becomes smoother and attributes are more likely to be independent.} At this time, we set $N$ to be a \textsf{sum} node and
call \textsf{FLAT-Offline} to model $\Pr_{T_i}(A_\N)$ with weight $w_i = {|T_i|}/{|T_\N|}$ for each $1 \leq i \leq n$ (lines~16--19).

\begin{figure}[t]
	\scriptsize
	\rule{\linewidth}{1pt}
	\leftline{~~~~\textbf{Algorithm} \textsf{FLAT-Offline$(A_{\N}, C_{\N}, T_{\N})$}}
	\vspace{-1em}
	\begin{algorithmic}[1]
		\IF{$C_{\N} = \emptyset$ }
		\STATE call \textsf{RDC}$(a, b, T_\N)$ for each pair of attributes $a, b \in A_\N$
		\STATE $H \gets \{a ,b | \textsf{RDC}(a, b, T_\N) \geq \tau_{h}\}$
		\STATE recursively enlarge $H \gets H \cup \{c |  \textsf{RDC}(a, c, T_\N) \geq \tau_{h} , a \in H, c \notin H\}$
		\IF{$H \neq \emptyset$}
		\STATE $O_{\N} \gets \textsf{factorize}$
		\STATE call $\textsf{FLAT-Offline}(A_\N - H, \emptyset, T_\N)$ on the left child of node $N$
		\STATE call $\textsf{FLAT-Offline}(H, A_\N - H, T_\N)$ on the right child of node $N$
		\ELSIF{$|A_\N| = 1$}
		\STATE $O_\N \gets \textsf{uni-leaf}$
		\STATE $\Pr_{\T_{\N}}(A_\N) \gets \textsf{Leaf-PDF}(A_{\N}, T_{\N})$
		\ELSIF{subsets $A_1, A_2, \dots, A_m$ are mutually indepedent}
		\STATE $O_{\N} \gets \textsf{product}$
		\STATE call $\textsf{FLAT-Offline}(A_i, \emptyset, T_\N[A_\N])$ on each child of $N$ for all $1 \leq i \leq m$
		\ELSE
		\STATE $O_{\N} \gets \textsf{sum}$
		\STATE $T_1, T_2, \dots, T_n \gets \textsf{Cluster}(T_\N, A_\N)$
		\STATE $w_i \gets {|T_i|}/{|T_\N|}$ for all $1 \leq i \leq n$
		\STATE call $\textsf{FLAT-Offline}(A_\N, \emptyset, T_i)$ with weight $w_i$ on each child of $N$ for $1 \leq i \leq n$
		\ENDIF
		\ELSE
		\STATE $ m \gets \max_{a \in \A_\N, c \in \C_\N} \textsf{RDC}(a, c, T_\N)$
		\IF{$m \leq \tau_{l}$}
		\STATE $O_\N \gets \textsf{multi-leaf}$
		\STATE $\Pr_{\T_{\N}}(A_\N) \gets \textsf{Leaf-PDF}(A_{\N}, T_{\N})$ 
		\STATE keep the range of $N$ in the attribute domain space
		\ELSE
		\STATE $O_{\N} \gets \textsf{split}$
		\STATE $ c \gets \arg\max_{a \in \A_\N, c \in \C_\N} \textsf{RDC}(a, c, T_\N)$
		\STATE divide $T_\N$ into $T_1, T_2, \dots, T_d$ by the range on attribute $c$
		\STATE call $\textsf{FLAT-Offline}(A_\N, C_\N, T_i)$ on each child of $N$ for $1 \leq i \leq d$
		\ENDIF
		\ENDIF
	\end{algorithmic}
	\vspace{-0.7em}
	\rule{\linewidth}{1pt}
	\vspace{-4em}
\end{figure}

\underline{\textit{3. Modeling conditional PDF (lines~21--30)}}: 
when $C_\N \neq \emptyset$, we try to model the conditional PDF $\Pr_{T_\N}(A_\N | C_\N)$. First, we compute pairwise correlations across all attributes in $A_\N$ and $C_\N$ (line~21). If $A_\N$ is independent of $C_\N$, $N$ is a \textsf{multi-leaf} node. \revise{We model the multivariate PDF $\Pr_{T_\N}(A_\N)$ using the piecewise regression technique~\cite{mczgee1970piecewise} and maintain its range in the attribute domain space (lines~23--25).}

Otherwise, we further split records in $T_\N$ (step~\textcircled{4} in Figure~\ref{fig: fspnstruc}). Probability computation requires $T_\N$ to be divided into grids in terms of $C_\N$. We apply a heuristic $d$-way partition method where $d$ is a hyper-parameter. We choose the attribute $c \in C_\N$ that maximizes the pairwise correlations between $A_\N$ and $C_\N$ (line~28). Intuitively, dividing the space by $c$ would largely break their correlations. We set $N$ to be a \textsf{split} node, evenly divide the range of $c$ on $T_\N$ into $d$ parts and get the clusters $T_1, T_2, \dots, T_d$ (line~29). After that, we call \textsf{FLAT-Offline} to model $\Pr_{T_i}(A_\N | C_\N)$ for each $1 \leq i \leq d$ (line~30).


\myskip
\noindent{\underline{\textbf{Complexity Analysis.}}}
Let $n$ be the number of nodes in the resulting \FSPN and $s$ be the number of \textsf{sum} nodes. On each inner node, we can sample a set of $r$ records from table $T$ to compute the RDC scores between attributes. The time cost of calling \textsf{RDC} is $O(r \log r)$, so the total time cost is $O(n {|A|}^{2} r \log r)$. On each \textsf{sum} node, we can also use the sampled records to compute the central points of the clusters and then assign each record to the nearest cluster. We denote the maximum iteration time in $k$-means as $t$. The total clustering time cost on all \textsf{sum} nodes is $O(s t k r)$. Besides, on each node, we need to scan all records in $T$ to assign them to the children (for inner nodes) or building the PDFs (for leaf nodes). The total scanning time cost is $O(n |T|)$. Therefore, the time complexity of \textsf{FLAT-Offline} is $O(n {|A|}^{2} r \log r + n |T| + s t k r)$. As $n$ is often small, it is efficient. By our testing, learning the structure of an \FSPN is faster than \textsf{SPN} and \textsf{DAR} to model the same joint PDF.




\subsection{Incremental Updates}
\label{sec: single-3}

When the table $T$ changes, we apply an incremental update method \textsf{FLAT-Update} to ensure the underlying \FSPN model can fit the new data. To attain high estimation accuracy while saving update cost,  we try to preserve the original \FSPN structure to the maximum extent while fine-tuning its parameters for better fitting. 

\revise{Let $\Delta T$ be the new data inserted  into \revise{(or deleted from)} $T$. We could traverse the \FSPN in a top-down manner to fit $T + \Delta T$ \revise{(or $T - \Delta T$)}. Specifically, 
for each \textsf{factorize} node $N$, since the conditional factorization is a lossless decomposition of the joint PDF, \revise{we directly propagate $\Delta T$ to its children}. For each \textsf{split} node, we propagate each record in $\Delta T$ to the corresponding child according to its splitting condition.}

\revise{
On each original \textsf{multi-leaf} node $L$, we recheck whether the conditional independence still holds after adding \revise{(or deleting)} some records. If so, we just update the parameters of its multivariate PDF by $\Delta T$. Otherwise, we reset it as a \textsf{split} node and run lines~28--30 of \textsf{FLAT-Offline} to further divide its domain space.}

\revise{For each \textsf{sum} node, we store the centroids of all clusters in structure construction. We could assign each record in $\Delta T$ to the nearest cluster (or remove each record from its original cluster), propagate it to that child and update the weight of each child accordingly.}

\revise{
For each \textsf{product} node, we also recheck whether the independence between attributes subset still holds after adding \revise{(or deleting)} some records. If not, we run lines~12--19 of \textsf{FLAT-Offline} to reconstruct the sub-structure of the \FSPNend. Otherwise, we directly pass $\Delta T$ to its children.
On each \textsf{uni-leaf} node, we update its parameters of the univariate PDF by $\Delta T$. Obviously, after updating, the generated \FSPN can accurately fit the PDF of $T + \Delta T$ \revise{(or $T - \Delta T$)}.}

\revise{Due to space limits, we put the pseudocode of \textsf{FLAT-Update} in Appendix~B.1 of the technical report~\cite{fullversion}.} It can run in the background of the DBMS. Note that, \textsf{FLAT-Update}  does not change the original \FSPN model when the data distribution keeps the same. In case of significant change of data or data schema changes, such as inserting or deleting attributes, the \FSPN could be rebuilt by calling \textsf{FLAT-Offline} in Section~4.2.

%


\begin{figure}[!h]
	\scriptsize
	\rule{\linewidth}{1pt}
	\leftline{~~~~\textbf{Algorithm} \textsf{FLAT-Multi$(D, Q)$}}
	\begin{algorithmic}[1]
		\STATE organize all tables in $D$ as a join tree $J$\% offline
		\FOR{each edge $(A, B) \in J$}
		\IF{\textsf{RDC}$(a, b) \geq \tau_{l}$ for any attribute $a$ of $A$ and $b$ of $B$}
		\STATE $A \gets \{A, B \}$
		\ENDIF
		\ENDFOR
		\FOR{each node $T$ in $J$ with attributes $A_{\T}$ of $\mathcal{T}$}
		\STATE add scattering coefficient columns $S_{\T}$ in $T$
		\STATE $\F_{\T} \gets \textsf{FLAT-Offline}(A_\T \cup S_\T, \emptyset, \mathcal{T})$
		\ENDFOR
		\STATE let $E = \{T_{1}, T_{2}, \dots, T_{d} \}$ denote all nodes in touched by $Q$ \% online
		\FOR{$i \gets 1$ to $d$}
		\STATE compute $p_i$ in Eq.~\eqref{eq: sqccorrect} by Technique~II
		\ENDFOR
		\RETURN $|\mathcal{E}| \cdot \prod_{i=1}^{d} p_i$
	\end{algorithmic}
	\rule{\linewidth}{1pt}
\end{figure}

\section{Multi-Table C\lowercase{ard}E\lowercase{st}  Method}
\label{sec: flat}

In this section, we discuss how to extend \FLAT algorithm to multi-table join queries. We first describe our approach on a high level, and then elaborate the key techniques in details.

\myskip
\noindent{\underline{\textbf{Main Idea.}}}
To avoid ambiguity, in the following, we use printed letters, such as $T, D$, to represent a set of tables, and calligraphic letters, such as $\mathcal{T, D}$, to represent the corresponding full outer join table. 
Given a database $D$, all information of $D$ is contained in $\mathcal{D}$. \textsf{DAR}-based approach~\cite{yang2020neurocard} builds a single large model on $\mathcal{D}$. It is easy to use and applicable to any type of joins between tables in $D$ but suffer from significant limitations.
First, no matter  how many tables are involved in a query, the entire model has to be used for probability computation, which may be inefficient. 
Second, the size of $\mathcal{D}$ grow rapidly w.r.t.~the number of tables in $D$, so its training cost is high even using
samples from $\mathcal{D}$.
Third, in case of data update of any table in $D$, the entire model needs to be retrained.

Another approach~\cite{hilprecht2019deepdb} builds a set of small models, where each captures the joint PDF of several tables $D' \subseteq D$. The joint PDF of attributes in $\mathcal{D}'$ (the full outer join table  of $D'$) is different from that in $\mathcal{D}$ since each record in $\mathcal{D}'$ can appear multiple times in $\mathcal{D}$. Therefore, the local model of $\mathcal{D}'$ needs to involve some additional columns to correct such PDF difference. When a query touches tables in multiple models, all local probabilities are corrected and merged together  to estimate the final cardinality. This approach is more efficient and flexible, but it only supports the primary-foreign key join. This is not practical as many-to-many joins are very common in query optimization (see Section~6.3 for examples on the benchmark workload).

\revise{
	To overcome their drawbacks, our approach absorbs the key ideas of~\cite{hilprecht2019deepdb} and also builds a set of small local models. However, we extend this method to be more general and applicable.
	First, we develop a new PDF correction paradigm, inspired by~\cite{hilprecht2019deepdb},  to support more types of joins, e.g., inner or outer and many-to-many (See the following Technique~I). Second, we specifically optimize the probability computation and correction process based on our \FSPN model (See Technique~II). Third, we develop incremental model updates method for data changes (See Technique~III).}

\begin{figure*}
\small
{
\begin{flushleft}
\begin{minipage}{0.08\linewidth}
	\centering
	\textbf{(a) Table $T_{\A}$} \\
	\begin{tabular}{|cc|}
	\hline
	\rowcolor{mygrey}
	$A_1$ & $A_2$ \\ \hline
	$b$ & 0 \\
	$c$ & 2 \\
	$b$ & 3 \\
	$c$ & 4 \\
	\hline
	\end{tabular}
\end{minipage}
\begin{minipage}{0.12\linewidth}
	\centering
	\textbf{(b) Table $T_{\B}$} \\
	\begin{tabular}{|ccc|}
		\hline
		\rowcolor{mygrey}
		$B_1$ & $B_2$ & $B_3$  \\ \hline
		1  & 0.3 & $M$ \\
		0  &  0.6 & $D$ \\
		3  & 0.4 & $D$ \\
		2  &  0.7& $M$ \\
		 4 & 0.5 & $K$ \\
		  3 & 0.2 & $K$ \\
		\hline
	\end{tabular}
\end{minipage}
\begin{minipage}{0.13\linewidth}
	\centering
	\textbf{(c) Table $T_{\C}$ of node $T_2$} \\
	\begin{tabular}{|cc|c|}
		\hline
		\rowcolor{mygrey}
		$C_1$ & $C_2$  & $S_{\T_2, \{\T_1, \T_2\} }$ \\ \hline
		$D$  & 0.2 & 2 \\
		$M$  & 0.7 & 2  \\
		$D$  & 0.8 & 2 \\
		$K$  & 0.9 & 2 \\
		\hline
	\end{tabular}
\end{minipage}
\begin{minipage}{0.45\linewidth}
	\centering
	\textbf{(d) Table $T_{\A} \fullouterjoin T_{\B}$ of node $T_1$} \\
	\begin{tabular}{|ccccc|cc|c|}
		\hline
		\rowcolor{mygrey}
		$A_1$ & $A_2$ & $B_1$ & $B_2$ & $B_3$ & $S_{\A, \B}$ & $S_{\B, \A}$
		& $S_{\T_1, \{\T_1, \T_2\} }$
		\\ \hline
		\textsf{null} & \textsf{null} & 1 & 0.3 & $M$ & 0 & 0 & 1 \\
		$b$ & 0 & 0 & 0.6 & $D$ & 1 & 1 & 2 \\
		$b$ & 3 & 3 & 0.4 & $D$ & 2 & 1 & 2 \\
		$c$ & 2 & 2 & 0.7 & $M$ & 1 & 1 & 1 \\
		$c$ & 4 & 4 & 0.5 & $K$ & 1 & 1 & 1 \\
		$b$ & 3 & 3 & 0.2 & $K$ & 2 & 1 & 1 \\
		\hline
	\end{tabular}
\end{minipage}
\hspace{-0.03\linewidth}
\begin{minipage}{0.205\linewidth}
	\centering
	\textbf{(e) Join Tree $J$ and Query $Q$} \\
	\includegraphics[width = \linewidth]{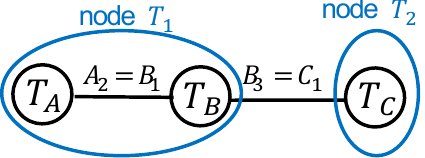}
	\fbox{\parbox{\textwidth}{
	$\boldsymbol{Q}$: \textsf{select count(*) from $T_\B$ full outer join $T_\C$ on $T_{\B.\B_3} \! = \! T_{\C.\C_1}$
	where  $T_{\B.\B_2}  \! \! > \! \! 0.5$ and $T_{\C.\C_2}  \! \! < \! \! 0.3$}
	}}
\end{minipage}
\end{flushleft}
\caption{Example databases and join query.}
\label{fig: flatexample}
}
\end{figure*}

\myskip
\noindent{\underline{\textbf{Algorithm Description.}}}
We present a high-level description of our approach in the \textsf{FLAT-Multi} algorithm, which takes a database $D$ and a query $Q$ as inputs. The main procedures are as follows:

\indent{\underline{\textit{1. Offline Construction (lines~1--7)}:}}
We first organize all tables in $D$ as a tree $J$ based on their joins. Initially, each node in $J$ is a table in $D$, and each edge in $J$ is a join between two tables. We do not consider self-join and circular joins in this paper. Based on $J$, we can partition all tables in $D$ into multiple groups such that: tables are highly correlated in the same group but weakly correlated in different groups.
Specifically, for each edge $(A, B)$ in $J$, we sample some records from $A \fullouterjoin B$, the outer join table, and examine the pairwise attribute correlation values between $A$ and $B$. If some correlation values are higher than a threshold, we learn the model on $A \fullouterjoin B$ together, so we merge $\{A, B\}$ to a single node. We repeat this process until no pair of nodes needs to be merged.
After that, the probability across different nodes can roughly be assumed as independent on their full outer join table. 

After the partition, each node $T$ in $J$ represents a set of one or more single tables. We add some scattering coefficient columns in its outer join table $\mathcal{T}$ for PDF correction. The details are explained in the following Technique~I. Then, we construct a \FSPN $\F_{\T}$ on $\mathcal{T}$ using \textsf{FLAT-Offline} in Section~4.2. If $\mathcal{T}$ is large, we do not explicitly materialize it. Instead, we draw some samples from $\mathcal{T}$ using the method in~\cite{zhao2018random} and train the \FSPN model on them.

Figure~\ref{fig: flatexample} depicts a example database with three tables. The join between $T_{\B}$ and $T_\C$ is a many-to-many join.  $T_{\A}$ and $T_{\B}$ are highly correlated so they are merged together into node $T_1$. Then, we build two \textsf{FSPN}s $\mathcal{F}_{\T_1}$ and $\mathcal{F}_{\T_2}$ on table $T_{\A} \fullouterjoin T_\B$  and $T_{\C}$, respectively.

\indent{\underline{\textit{2. Online Processing (lines~8--12):}}}
Let $E = \{T_{1}, T_{2}, \dots, T_{d} \}$ denote all nodes in $J$ touched by the query $Q$ and $Q_i$ be the sub-query on $T_i$. By our assumption, the probability of each $Q_i$ is independent on the table $\mathcal{E} = \mathcal{T}_1 \fullouterjoin \mathcal{T}_2 \fullouterjoin \dots \fullouterjoin \mathcal{T}_d$. 
We can efficiently correct the probability from the local model $\F_{\T_i}$ on $\mathcal{T}_i$ to $\mathcal{E}$ by a new paradigm. Finally, we multiply all probabilities to get the final result.

\smallskip

\myskip
\noindent{\underline{\textbf{Technique~I: Probability Correction Method.}}}
We need to correct the probability to account for the effects of joining from two aspects. We elaborate the details with the example query $Q$ in Figure~\ref{fig: flatexample}(e). $Q$ is divided into two sub-queries: $Q_1$ ($T_{\B.\B_2}  \! > \!  0.5$ on node $T_1$) and $Q_2$ ($T_{\C.\C_2}  <  0.3$ on node $T_2$). 
First, on node $T_1$, the \FSPN $\mathcal{F}_{\T_1}$ is built on table $T_{\A} \fullouterjoin T_\B$ instead of table $T_{\B}$ individually. As each record in $T_{\B}$ can occur multiple times in $T_{\A} \fullouterjoin T_\B$, the probability obtained by $\mathcal{F}_{\T_1}$ needs to be \emph{down-scaled} to remove the effects of $T_\A$.
Second, the probability obtained on node $T_2$ is defined on table $T_\C$ individually but not on $T_{\B} \fullouterjoin T_\C$. Therefore, the probability of $Q_2$ (and also $Q_1$) needs to be \emph{up-scaled} to add the effects of joining. 

\revise{The above corrections are achieved by adding extra columns in table $\mathcal{T}_i$ of each node $T_i$. These columns track the number of times that a record in a single table $A$ appears in $\mathcal{T}_i$, i.e., the scattering effect. Previous works~\cite{hilprecht2019deepdb,yang2020neurocard} add columns to process the scattering effects of each join in only one side. However, our solution considers the scattering effects on two sides of each join. It is more practical by supporting more join types in one framework, and more general by processing down-scale and up-scale effects at the same time.
}

For each pair of joined tables $(A, B)$ in a node $T_i$, we add two additional attributes $S_{\A, \B}$ and $S_{\B, \A}$ in $\mathcal{T}_i$. $S_{\A, \B}$ indicates how many records in $B$ can join with this record in $A$ and vice versa. We call such $S_{\A, \B}$  \emph{scattering coefficient}. In Figure~\ref{fig: flatexample}(d), we add two columns $S_{\A, \B}$ and $S_{\B, \A}$ in the table $T_{\A} \fullouterjoin T_{\B}$ of $T_1$. These columns are be used to down-scale the effects of untouched tables inside each node.

Similarly, for up-scale correction, we can regard node $T_i$ as the root of the join tree $J$. For each distinct sub-tree of $J$ rooted at $T_i$ containing nodes $E' = \{T_1', T_2', \dots, T_d'\}$, we add a column $S_{\T_{i}, E'}$ in table $\mathcal{T}_i$ indicating the scattering coefficient of each record in $\mathcal{T}_{i}$ to the outer join table $\mathcal{E}' = \mathcal{T}'_1 \fullouterjoin \mathcal{T}'_2 \fullouterjoin \dots \fullouterjoin \mathcal{T}'_d$. For the node $T_2$ in Figure~\ref{fig: flatexample}(c), we add the column $S_{\T_2, \{\T_1, \T_2\} }$ indicating the scattering coefficient of each record in $T_{\C}$ when joining with $T_{\A} \fullouterjoin T_\B$. 
The method to compute the values of these scattering coefficient columns has been proposed in~\cite{zhao2018random}. Briefly speaking, we can obtain the values of $S_{\T_{i}, E'}$ by recursively aggregating over all sub-trees rooted at $T_i$'s children. Using dynamic programming, the time cost of computing scatter coefficient values over all nodes is linear w.r.t.~table size. 

As all tables form a join tree, the number of added scattering columns in each node is linear w.r.t.~its number of tables.
In each node $T_i$, all scattering coefficient columns are learned together with other attributes when constructing the \FSPN $\mathcal{F}_{\T_i}$. 

We can estimate the cardinality by the following lemma. We put the detailed correctness proof in Appendix~C of the technical report~\cite{fullversion}.  In a high order, for each record with down-scale value $s$ and up-scale value $e$, we correct its probability satisfying $Q_i$ by a factor of $e/s$.
We set $e$ or $s$ to $1$ if it is $0$ since records with zero scattering coefficient also occur once in the full outer join table. 

\textbf{\textit{Lemma~2}}
\textit{
Given a query $Q$, let $E = \{T_{1}, T_{2}, \dots, T_{d} \}$ denote all nodes in $J$ touched by $Q$. On each node $T_i$, let $S = \{ S_{\A_{1}, \B_{1}}, S_{\A_{2}, \B_{2}}, \dots, \break S_{\A_{n}, \B_{n}}\}$, where each $(A_j, B_j)$ is a distinct join such that $B_j$ is not in $Q$. Let $s = (s_1, s_2, \dots, s_n)$ where $S_{\A_{j}, \B_{j}} = s_j  \in \mathbb{N}$ for all $1 \leq i \leq n$ denote an assignment to $S$ and $\text{dlm}(s) = \prod_{j=1}^{n} \max\{s_j, 1\}$. Let}
\begin{equation}
\label{eq: sqccorrect}
p_i \! = \frac{|\mathcal{T}_i|}{|\mathcal{E}|} \! \cdot \! 
\sum\limits_{s, e}  \left( \Pr\nolimits_{\mathcal{T}_i}(Q_i  \wedge S \! = \! s \wedge S_{\T_{i}, \E} \! = \! e) \cdot \frac{\max\{e, 1\}}{\text{dlm}(s)} \right).
\end{equation}
\textit{Then, the cardinality of $Q$ is $|\mathcal{E}| \cdot \prod_{i = 1}^{d} p_i$.}

Consider again query $Q$ in Figure~\ref{fig: flatexample}(e). For the sub-query $Q_1$ on node $T_1$, we need to down-scale by $S_{\B, \A}$ and up-scale by 
$S_{\T_1, \{\T_1, \T_2\}}$. By Eq.~\eqref{eq: sqccorrect}, we have $p_1 = (1*2 + 1 + 1)/8 = 1/2$. Similarly, we have $p_2 = 1/4$ for sub-query $Q_2$, so the final cardinality of $Q$ is $8 * (1/8) = 1$.

As a remark,  if two tables $A$ and $B$ are inner joined in $Q$, we can add the constraint $S_{\A, \B} \! > \! 0$ and $S_{\B, \A} \! > \! 0$ (or $S_{\A, \E} \! > \! 0$ and $S_{\B, \E} \! > \! 0$ if $A$ and $B$ in different nodes) in Eq.~\eqref{eq: sqccorrect} to remove all records in $A$ or $B$ that have no matches. Similarly, we only add $S_{\B, \A} > 0$ or $S_{\A, \B} > 0$ to $Q$ for left and right join, respectively.

\myskip
\noindent{\underline{\textbf{Technique~II: Fast Probability Computation:}}
Notice  that, the value $p_i$ in Eq.~\eqref{eq: sqccorrect} involves summing over the probabilities of each assignment to the down-scale value $s$ and up-scale value $e$. If we directly obtain all these probabilities, the time cost is very high.
Instead, we present an optimized method to compute $p_i$, which only requires a \emph{single} traversal on the underlying \FSPN model.

Specifically, on any node $T$ in the join tree, let $S_{\T}$ and $A_{\T}$ denote the scattering coefficient and attribute columns in $\mathcal{T}$, respectively. When constructing the \FSPN $\F_{\T}$, we first use a \textsf{factorize} root node to split the joint PDF $\Pr_{\mathcal{T}}(S_{\T}, A_{\T})$ into $\Pr_{\mathcal{T}}(A_{\T})$ on the left child $LC$ and $\Pr_{\mathcal{T}}(S_{\T} | A_{\T})$ on the right child $RC$. Each leaf node $L$ of $RC$ models a PDF of $S_{\T}$. By \FSPNend's semantic, the probabilities of any query $Q$ on $A_{\T}$ and $S_{\T}$ are independent on each $L$. Then, we have
\begin{equation}
\label{eq: tcqcomplexfast}
\small
\begin{split}
\Pr'\nolimits_{\mathcal{T}} (Q)  \! & = \!\! \sum_{\L}  \left( \Pr\nolimits_{\L}(A_\T) \! \cdot \! 
\sum_{s, e} \left( \Pr\nolimits_{\L}(S = s \wedge S_{\T, \E} = e) \cdot \frac{\max\{e, 1\}}{\text{dlm}(s)} \right) \right) \\ 
& = \!\!  \sum_{\L}  \left( \Pr\nolimits_{\L}(A_\T) \! \cdot \mathbb{E} \left[\frac{\max\{e, 1\}}{\text{dlm}(s)} \right] \! \right).
\end{split}
\end{equation}

For the left part, the probability $\Pr_{\L}(A_\T)$ could be computed with the \FSPN rooted at node $LC$ using the method in Section~4.1. For the right part, it is a \emph{fixed} expected value of $\max\{e, 1\} / \text{dlm}(s)$ of $S_{\T}$. 
Therefore, we can pre-compute the expected value for each possible $S, S_{\T, \E} \subseteq S_\T$ on each leaf $L$. After that, each $p_i$ in Eq.~\eqref{eq: sqccorrect} could be obtained by traversing the \FSPN $\mathcal{F}_{\T_i}$ \emph{only once}. \revise{By our empirical analysis in Section~4.1, the \CE time cost for multi-table queries is also near linear w.r.t.~the number of nodes in \textsf{FSPN}s.}

\myskip
\noindent{\underline{\textbf{\color{black} Technique~III: Incremental Updates.}}
{\color{black} Next, we introduce how to update the 
underlying \FSPN models in multi-table cases. We put the pseudocode of our algorithm \textsf{FLAT-Update-Multi} in Appendix~B.2~\cite{fullversion} and describe the procedures as follows.

First, we consider the case of inserting some records $\Delta C$ in a table $C$ of the node $T$. It affects $\mathcal{T}$ in three aspects:
1) each record in $\Delta C$ can join with other tables in $T$. We use $\Delta \mathcal{T_{+}}$ to denote all new records inserted into $\mathcal{T}$;
2) each record in $\mathcal{T}$, which does not find a match in table $C$ (\textsf{null}) but can join with the new records in $\Delta C$, needs to be removed. We denote them as $\Delta \mathcal{T_{-}}$;
and 3) the scattering coefficient of each record in $\mathcal{T}$, which can join with new records in $\Delta C$, needs to be enlarged. We denote these records as $\Delta \mathcal{T_{*}}$.
We can directly join $\Delta C$ with $\mathcal{T}$ to identify
$\Delta \mathcal{T_{+}}$, $\Delta \mathcal{T_{-}}$ and $\Delta \mathcal{T_{*}}$ accordingly.
}

\revise{
Next, we describe how to incrementally update the \FSPN $\F_{\T}$ built by Technique~II. Recall that the root node $N$ of $\F_{\T}$ is a \textsf{factorize} node separating attributes and scattering coefficient columns, which enables fast incremental update. The left child $LC$ of $N$ models $\Pr_{\mathcal{T}}(A_{\T})$ on all attribute columns. We could update it to fit the data $\mathcal{T} + \Delta \mathcal{T_{+}} - \Delta \mathcal{T_{-}}$ by directly calling the \textsf{FLAT-Update} method in Section~4.3. The right child $RC$ of $N$ models $\Pr_{\mathcal{T}}(S_{\T} | A_{\T})$ on all scattering coefficients columns.
Each multi-leaf $L$ of $RC$ only stores some expected values of $S_{\T}$ defined by Eq.~\eqref{eq: tcqcomplexfast}. We can pre-build a hash table on the probability of each assignment $s$ of $S_{\T}$. Then,
based on the changes of scattering columns in $\Delta \mathcal{T_{+}}$, $\Delta \mathcal{T_{-}}$ and $\Delta \mathcal{T_{*}}$, we can incrementally update all expected values.
}

\revise{
Finally,  as $\mathcal{T}$ changes, we need to propagate the effects to other nodes $T'$ to update all scattering columns $S_{\T', \E}$. For efficiency, it can run in the background asynchronously. Specifically, after each time interval such as one day, we scan all tables and recompute the scattering coefficients using the method in~\cite{zhao2018random}. Then we incrementally update the expected values stored in \FSPN $\mathcal{F}_{\T'}$. 
}

\revise{
For the case of deleting some records $\Delta C$ in a table $C$ of the node $T$, the updating could be done in a very similar way. At this time,
we obtain $\Delta \mathcal{T_{-}}$ containing all removed tuples joining with $\Delta C$ previously, $\Delta \mathcal{T_{+}}$ containing all added tuples having no matches in table $C$ and $\Delta \mathcal{T_{*}}$ containing all original records whose scattering coefficients are reduced. Then we update the \FSPN $\F_{\T}$ and $\F_{\T'}$ of other nodes $T'$ in the same way as the insertion case. 
Notice that, the data insertion and deletion can also be done simultaneously as long as we maintain the proper set of records $\Delta \mathcal{T_{+}}$, $\Delta \mathcal{T_{-}}$ and $\Delta \mathcal{T_{*}}$. 
In the complex case of creating new tables or deleting existing tables in the database, the model could be retrained offline.
}


\section{Evaluation Results}
\label{sec: exp}
We have conducted extensive experiments to demonstrate the superiority of our proposed \FLAT algorithm. We first introduce the experimental settings, and then report the evaluation results of \CE algorithms on the single table and multi-table cases in Section~6.1 and~6.2, respectively. Section~6.3 reports the effects of updates.
Finally, in Section~6.4, we integrate \FLAT into the query optimizer of Postgres~\cite{psql2020} and evaluate the end-to-end query optimization performance.

\myskip
\noindent{\underline{\textbf{Baselines.}}}
We compare \FLAT with a variety of representative \CE  algorithms, including:\\
\indent 1) \textsf{Histogram}: the simplest \revise{\textsf{1-D histogram}} based \CE method widely used in DBMS such as SQL Server~\cite{sqlserver2019} and Postgres~\cite{psql2020}. \\
\indent 2) \textsf{Naru}: a \textsf{DAR} based algorithm proposed in~\cite{yang2019deep}. We adopt the authors' source code from~\cite{yang2019naru} with the var-skip speeding up technique~\cite{liang2020variable}. It utilizes a DNN with 5 hidden layers (512, 256, 512, 128, 1024 neuron units) to approximate the PDFs. The sampling size is set to $2, 000$ as the authors' default. We do not compare with the similar method  in~\cite{hasan2019multi}, since their performance is close.\\
\indent 3)
\textsf{NeuroCard}~\cite{yang2020neurocard}: an extension of \textsf{Naru} onto the multi-table case. We also adopt the authors' source code from~\cite{yang2020sb} and set the sampling size to $8, 000$ as the authors' default. 
\\
\indent 4) \textsf{BN}: a Bayesian network based algorithm. We use the Chow-Liu Tree~\cite{chow1968approximating, dasfaa2019} based implementation to build the \textsf{BN} structure, since its performance is better than others~\cite{getoor2001selectivity, tzoumas2011lightweight}. \\
\indent 5) \textsf{DeepDB}: a \textsf{SPN} based algorithm proposed in~\cite{hilprecht2019deepdb}. We adopt the authors' source code from~\cite{hilp2019deepdb} and apply the same hyper-parameters, which set the RDC independence threshold to $0.3$ and split each node with at least $1\%$ of the input data. \\
\indent 6) \revise{\textsf{SPN-Multi}:}
\revise{a simple extension of \textsf{SPN} with multivariate leaf nodes. It maintains a \textsf{multi-leaf} node if the data volume is below $1\%$ and attributes are still not independent.\\}
\indent 7) \revise{\textsf{MaxDiff}: } 
\revise{a representative M-D histogram based method~\cite{poosala1997selectivity}. We use the implementation provided in the source code repository of ~\cite{yang2019naru}. We do not compare with the improved methods DBHist~\cite{deshpande2001independence}, GenHist~\cite{gunopulos2000approximating} and VIHist~\cite{wang2003multi} are they are not open-sourced.}\\
\indent 8) \textsf{Sample}:  the method uniformly samples a number of records to estimate the cardinality. We set the sampling size to $1\%$ of the dataset. It is used in DBMS such as MySQL~\cite{mysql2020} and MariaDB~\cite{mdb2020}. We do not compare with other method such as IBJS~\cite{leis2017cardinality} since their performance has been verified to be less competitive~\cite{yang2019naru, yang2020neurocard, hilprecht2019deepdb}.
\\
\indent 9) \textsf{KDE}:  kernel density estimator based method for \CEend. We have implemented it using the scikit-learn module~\cite{kde2020}. \\
\indent 10) \textsf{MSCN}: a state-of-the-art query-driven \CE algorithm described in~\cite{kipf2018learned}. For each dataset, we train it with $10^{5}$ queries generated in the same way as the workload.
}

Regarding \FLAT hyper-parameters as described in Section~4.2, we set the RDC threshold $\tau_{l} = 0.3$ and $\tau_{h} = 0.7$ for filtering independent and highly correlated attributes, respectively, and set $d = 2$ for $d$-way partition of records. 
Similar to \textsf{DeepDB}, we also do not split a node when it contains less than $1\%$ of the input data. \revise{
The sensitivity analysis of hyper-parameters are put in Appendix~D~\cite{fullversion}.}

\myskip
\noindent{\underline{\textbf{Evaluation Metrics.}}}
Based on our discussion in Section~1, we concentrate on examining three key metrics: estimation accuracy, time efficiency and storage overhead. For estimation accuracy, we adopt the widely used q-error metric~\cite{yang2019deep,hilprecht2019deepdb,hasan2019multi,kipf2018learned,leis2018query, leis2015good} defined as the larger value of$\Card(T, Q)/\widehat{\Card}(T, Q)$ and $\widehat{\Card}(T, Q)/\Card(T, Q)$, so its optimal value $1$.
We report the whole q-error distribution ($50\%$, $90\%$, $95\%$, $99\%$ and $100\%$ quantile) of each workload. 
For time efficiency, we report the estimation latency and model training time. For storage overhead, we report the model size.

\myskip
\noindent{\underline{\textbf{Environment.}}}
All above algorithms have been implemented in Python. All experiments are performed on a CentOS Server with an Intel Xeon Platinum 8163 2.50GHz CPU having 64 cores, 128GB DDR4 main memory and 1TB SSD.

\subsection{Single Table Evaluation Results}
\label{sec: exp-1}

We use two single table datasets: 1) GAS is real-world gas sensing data obtained from the UCI dataset~\cite{Gas2020} and contains 3,843,159 records. We extract the most informative 8 columns (\textsl{Time}, \textsl{Humidity}, \textsl{Temperature}, \textsl{Flow\_rate}, \textsl{Heater\_voltage}, \textsl{R1}, \textsl{R5} and \textsl{R7});
and 2) DMV~\cite{DMV2020} is a real-world vehicle registration information dataset and contains 11,591,877 tuples. We use the same 11 columns as~\cite{yang2019naru}. 

For each dataset, we generate a workload containing $10^{5}$ randomly generated queries. For each query, we use a probability of $0.5$ to decide whether an attribute should be contained. As stated in Section~2, the domain of each attribute $A$ is mapped into an interval, so we uniformly sample two values $l$ and $h$  from the interval such that $l \leq h$ and set $A \in [l, h]$. 

\myskip
\noindent{\underline{\textbf{Estimation Accuracy.}}}
Table~\ref{tab: exp-acc-qerror-single} reports the q-error distribution for different \CE algorithms. 
\revise{As main take-away, their accuracy can be ranked as $\FLAT \! \approx \! \textsf{Naru} \! \! \approx \! \textsf{SPN-Multi} \! > \! \textsf{BN} \! > \! \textsf{DeepDB} \! >>\! \textsf{Sample}/\textsf{MSCN} >> \textsf{KDE} >> \textsf{MaxDiff}/\textsf{Histogram}$.} The details are as follows:

\revise{
1) Overall, \FLAT's estimation accuracy is very high. On both datasets, the median q-error (1.001 and 1.002) is very close to $1$, the optimal value. On GAS, \FLAT attains the highest accuracy. The accuracy of \textsf{Naru} and \textsf{SPN-Multi} is comparable to \FLATend, which is marginally better than \FLAT on DMV. The high accuracy of \textsf{Naru} and stems from its \textsf{AR} based decomposition and the large DNN representing the PDFs. \textsf{SPN-Multi} achieves high accuracy as it models the PDFs of attributes without independence assumption.}

2) The accuracy \textsf{BN} and \textsf{DeepDB} is worse than \FLATend. At the $95\%$ quantile, \FLAT outperforms \textsf{BN} by $3.6 \times$ and \textsf{DeepDB} by $71 \times$ on GAS. The error of \textsf{BN} mainly arises from its approximate structure construction. \textsf{DeepDB} appears to fail at splitting highly correlated attributes. Thus, it causes relatively large estimation errors for queries involving these attributes.

3) The accuracy of \textsf{MSCN} and \textsf{Sample} appears unstable. \FLAT outperforms \textsf{MSCN} by $109 \times$ and $1.8 \times$ on GAS and DMV, respectively. As \textsf{MSCN} is query-driven, its accuracy relies on if the workload is ``similar'' to the training samples. Whereas, \FLAT outperforms \textsf{Sample} by $4.5 \times$ and $56 \times$ on GAS and DMV, respectively as the sampling space of DMV is much larger than GAS.

4) \revise{\FLAT largely outperforms \textsf{Histogram}, \textsf{MaxDiff} and \textsf{KDE} since \textsf{Histogram} and \textsf{MaxDiff} makes coarse-grained 
independence assumption and \textsf{KDE} may not well characterize high-dimensional data by tuning a good bandwidth for kernel functions~\cite{kiefer2017estimating}.}

\begin{table}
	\caption{Performance of \CE algorithms on single table.}
	\resizebox{\columnwidth}{!}{
		\begin{tabular}{c|c|ccccc|c|c}
			\hline
			\rowcolor{mygrey}
			& & & & & & & & \large \bf Training \\
			\rowcolor{mygrey}
			\multirow{-2}{*}{\large \bf Dataset} & \multirow{-2}{*}{\large \bf Algorithm} &
			\multirow{-2}{*}{\large \bf 50\%} & \multirow{-2}{*}{\large \bf 90\%} & \multirow{-2}{*}{\large \bf 95\%} & \multirow{-2}{*}{\large \bf 99\%} & \multirow{-2}{*}{\large \bf Max} & \multirow{-2}{*}{\large \bf Size (KB) }& \large \bf Time (Min) \\ 
			\hline
			\multirow{9}{*}{GAS} & \textsf{Histogram} & 2.732 & 53.60 & 163.0 & $2\cdot10^6$ &$3\cdot10^7$ & \large \textbf{34} & \large \textbf{1.3} \\
			& \textsf{Naru} & 1.007 & 1.145 & 1.340 & 2.960 &16.50 & 6, 365 & 216\\
			& \textsf{BN} & 1.011 & 1.208 & 1.550 & 4.780 & 36.80 & 108 & 8.2\\
			& \textsf{DeepDB} & 1.039 & 1.765 & 2.230 &95.12 &619.2 & 218 & 54\\
			& \revise{\textsf{SPN-Multi}} & 1.005 & 1.169 & 1.289 & 1.461 & 3.702 & 31,253 & 62 \\
			& \textsf{MaxDiff} & 2.211 & 86.7 & 196.0 & $3\cdot10^4$ &$8\cdot10^5$ & $3\cdot10^5$ & 310\\
			& \textsf{Sample} & 1.046 & 1.625 &2.064 & 6.017 & 3, 410 &- & -\\ 
			& \textsf{KDE} &3.307 & 5.469 &6.742 & 471.0 & $2\cdot10^4$ &- & 27\\
			& \textsf{MSCN} & 2.610 & 68.47 & 129.0 & $1\cdot10^5$ & $7\cdot10^5$ &2, 663 & 662\\
			&  \textbf{\textsf{FLAT} (Ours)} & \large \textbf{1.001} &  \large \textbf{1.127} &  \large \textbf{1.183} &  \large \textbf{1.325} & \large \textbf{3.178} &  198 & 19 \\
			\hline
			\multirow{9}{*}{DMV} & \textsf{Histogram} & 1.184 & 2.541 & 41.72 & 710.0 & $2\cdot10^5$ & \large \textbf{24} & \large \textbf{1.6}\\
			& \textsf{Naru} & 1.006 & 1.184 & 1.368& \large \textbf{6.907} & \large \textbf{49.03} & 7, 564 & 146\\
			& \textsf{BN} & 1.003 & 1.264 & 1.818 & 9.800 & 176.0 & 59 & 5.4\\
			& \textsf{DeepDB} & 1.005 & 1.574 &2.604 &27.90 & 534.0 &247 & 48\\
			& \revise{\textsf{SPN-Multi}} & 1.004 & \large \textbf{1.163} & \large \textbf{1.347} & 7.225 & 58.37 & 53,267 & 53 \\
			& \textsf{MaxDiff} & 1.802 & 6.304 & 28.81 & 4, 320 & $3\cdot10^4$ & $7\cdot10^5$ & 249\\
			& \textsf{Sample} & 1.122 & 1.619 & 9.010 & 551.0 & 7, 077 & - &-\\ 
			& \textsf{KDE} & 3.493 & 15.07 & 104.0 & 589.0 & $5\cdot10^4$ & - & 48\\
			& \textsf{MSCN} & 1.215 & 2.612 & 4.420 & 17.90 & 1, 192 & 2, 566 & 744\\
			&  \textbf{\textsf{FLAT} (Ours)} &  \large \textbf{1.002} &  1.255 & 1.795 &  9.805 & 76.50 & 53 & 2.4\\
			\hline
	\end{tabular}}
	\label{tab: exp-acc-qerror-single}
\end{table}

\myskip
\noindent{\underline{\textbf{Estimation Latency.}}}
Figure~\ref{fig: exp-latency} reports the average latency of all \CE methods. Since only \textsf{MSCN} and \textsf{Naru} provide the implementation optimized for GPUs, we compare all \CE methods on CPUs
for fairness. We provide the comparison results on GPUs in Appendix~E.1~\cite{fullversion}. 
\revise{In summary, their speed on CPUs can be ranked as $\textsf{Histogram} \approx \FLAT > \textsf{MSCN}  > \textsf{SPN-Multi} /\textsf{DeepDB} > \textsf{KDE} /\textsf{Sample}  >\!> \text{Others}$.} 
The details are as follows:

1) \textsf{Histogram} runs the fastest, it requires around $0.1ms$ for each query. \FLAT is close with a latency around $0.2ms$ and $0.5ms$ on DMV and GAS, respectively. Both are much faster than all other methods. This can be credited to the \FSPN model used in \FLAT being both compact and easy to traverse for probability computation. \textsf{MSCN} is also fast since it only requires a forward pass over DNNs.

2) \revise{\textsf{DeepDB}, \textsf{SPN-Multi}, \textsf{KDE} and \textsf{Sample} need up to $10ms$ for each query. \FLAT is $1$--$2$ orders of magnitude faster than them because the \FSPN model used in our \FLAT is more compact than the \textsf{SPN} model in \textsf{DeepDB} \revise{and \textsf{SPN-Multi}}.} In addition, \textsf{KDE} and \textsf{Sample} need to examine large amount of samples, thus less efficient.

3) \textsf{MaxDiff}, \textsf{BN} and \textsf{Naru} need $10$--$100ms$ for each query. \FLAT is $2$--$3$ orders of magnitude faster than them, e.g., $213 \times$ and $599 \times$ faster than \textsf{Naru} on GAS and DMV, respectively. The time cost of \textsf{MaxDiff} is spent on decompressing the joint PDF. The inference on \textsf{BN} is NP-hard and hence inefficient. \textsf{Naru} requires repeated sampling for range querie so it is computationally demanding. 

\begin{figure}[t]
	\includegraphics[width=0.85\columnwidth]{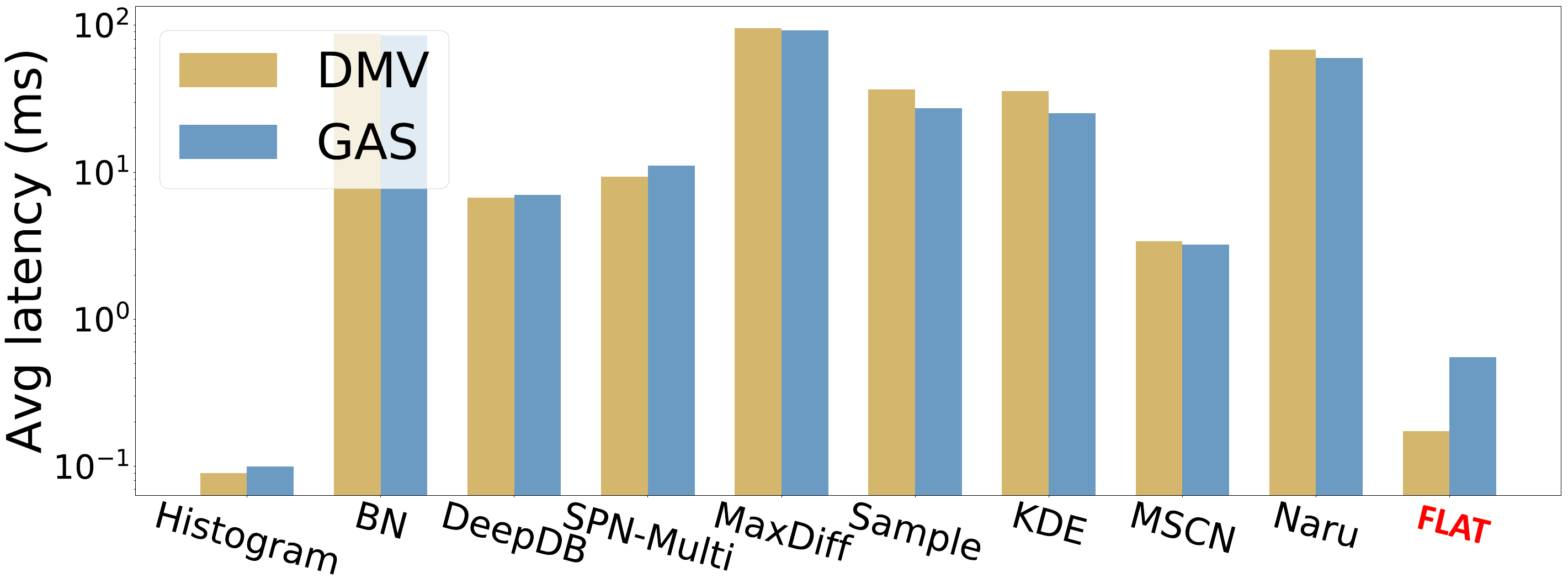}
	\caption{\revise{Estimation latency of \CE algorithms.}}
	\label{fig: exp-latency}
\end{figure}

\myskip
\noindent{\underline{\textbf{Model Training Time.}}}
As shown in the last column in Table~\ref{tab: exp-acc-qerror-single}, \FLAT is very efficient in training. 
Specifically, on DMV, \FLAT is $61 \times$ and $20 \times$ faster than \textsf{Naru} and \textsf{DeepDB} in training. This is due to the structure of \FSPN is much smaller than \textsf{SPN}, and our training process does not require iterative gradient updates as required for SGD-based  training of DNNs~\cite{Bottou2012Stochastic}.

\myskip
\noindent{\underline{\textbf{Storage Overhead.}}}
Storage costs are given in Table~\ref{tab: exp-acc-qerror-single}. The storage cost of \textsf{Histogram} and \textsf{BN} is proportional to the attribute number so they require the smallest storage. \FLAT is also very small requiring about $2\times$ of \textsf{Histogram}. \textsf{DeepDB} requires more storage space than \FLAT since the learned \textsf{SPN} has more nodes. They consume $10$--$100$KB of storage. \textsf{MSCN} and \textsf{Naru} consume several MB since they store  large DNN models. 
\revise{\textsf{SPN-Multi} requires tens of MB as it needs to maintain the \textsf{multi-leaf} nodes on not highly correlated attributes, as we discussed in Section~3.} The storage cost of \textsf{MaxDiff} is the highest since it stores the compressed joint PDF. 

\myskip
\noindent{\underline{\textbf{Model Node Number.}}}
\revise{To give more details, we also compare the number of nodes (or neurons) in \textsf{DeepDB}, \textsf{SPN-Multi} and \textsf{Naru}.} The 5-layer DNN in \textsf{Naru} is fully connected and contains $2,432$ neurons. The \textsf{SPN} used in \textsf{DeepDB} contains $873$ and $823$ nodes on GAS and DMV, respectively. \revise{\textsf{SPN-Multi} contains $825$ and $787$ nodes on GAS and DMV, respectively.} Whereas, the \FSPN in \FLAT only uses $210$ and $20$ nodes on GAS and DMV, respectively. \revise{\FSPN uses $21 \times$, $7.4 \times$ and $7 \times$ less nodes than DNN, \textsf{SPN} and \textsf{SPN-Multi} to model the same joint PDF}.

\myskip
\noindent\revise{{\underline{\textbf{Stability.}}}}
\revise{We also examine \FLAT on synthetic datasets. The results in Appendix~E.2 show that \FLAT is stable to varied correlations and distributions and relatively robust to varied domain size.
}

\subsection{Multi-Table Evaluation Results}
\label{sec: exp-2}

We evaluate the \CE algorithms for the multi-table case on the IMDB benchmark dataset. It has been extensively used in prior work~\cite{leis2018query, leis2015good, yang2020neurocard, hilprecht2019deepdb} for cardinality estimation. We use the provided \textit{JOB-light} query workload with 70 queries and create another more complex and comprehensive workload \textit{JOB-ours} with $1,500$ queries. 

\textit{JOB-light}'s schema contains six tables (\textsl{title}, \textsl{cast\_info}, \textsl{movie\_info}, \textsl{movie\_companies}, \textsl{movie\_keyword}, \textsl{movie\_info\_idx}) where all other tables can only join with \textsl{title}.
Each \textit{JOB-light} query involves $3$--$6$ tables with $1$--$4$ filtering predicates on all attributes.
\textit{JOB-ours} uses the same schema as \textit{JOB-light} but each query is a range query using $4$--$6$ tables and $2$--$7$ filtering  predicates. The predicate of each attribute is set in the same way as on single table. 
Figure~\ref{fig: exp-card-dist} illustrates the true cardinality distribution of the two workloads. The scope of cardinality for \textit{JOB-ours} is wider than \textit{JOB-light}. 
Note that, the model of each \CE method is the same for the two workloads.
\revise{As the attributes are highly correlated on IMDB, the model size of \textsf{SPN-Multi} exceeds our memory limit, so we can not evaluate it.
}

\myskip
\noindent{\underline{\textbf{Results on \textit{JOB-light}.}}}
Table~\ref{tab: exp-acc-qerror-multi} reports the q-error and storage cost of \CE methods on the \textit{JOB-light} workload. We observe that:

1) The accuracy of \FLAT is the highest among all algorithms. \textsf{NeuroCard} is only a bit better w.r.t the maximum q-error, which reflects only one query in the workload. At the $95\%$ quantile, \FLAT outperforms \textsf{NeuroCard}  by $2.6 \times$, \textsf{BN} by $33 \times$, \textsf{DeepDB} by $1.7 \times$ and \textsf{MSCN} by $43 \times$. The reasons have been explained in Section~6.1.

2) In terms of  storage size, \textsf{Histogram} and \textsf{BN} are still the smallest and \textsf{MaxDiff} is still the largest. \FLAT's space cost is $3.3$MB, which is 
$10.8 \times$ and $2.1 \times$ less than \textsf{DeepDB} and \textsf{NeuroCard}, respectively. In comparison with the single table case, \FLAT's space cost is relatively large. This is because for the multi-table case, \FSPN needs to process more attributes---the scattering coefficients columns and materialize some values for fast probability computation. However, it is still 
reasonable and affordable for modern DBMS.

\begin{figure}[t]
	\includegraphics[width = 0.85\columnwidth]{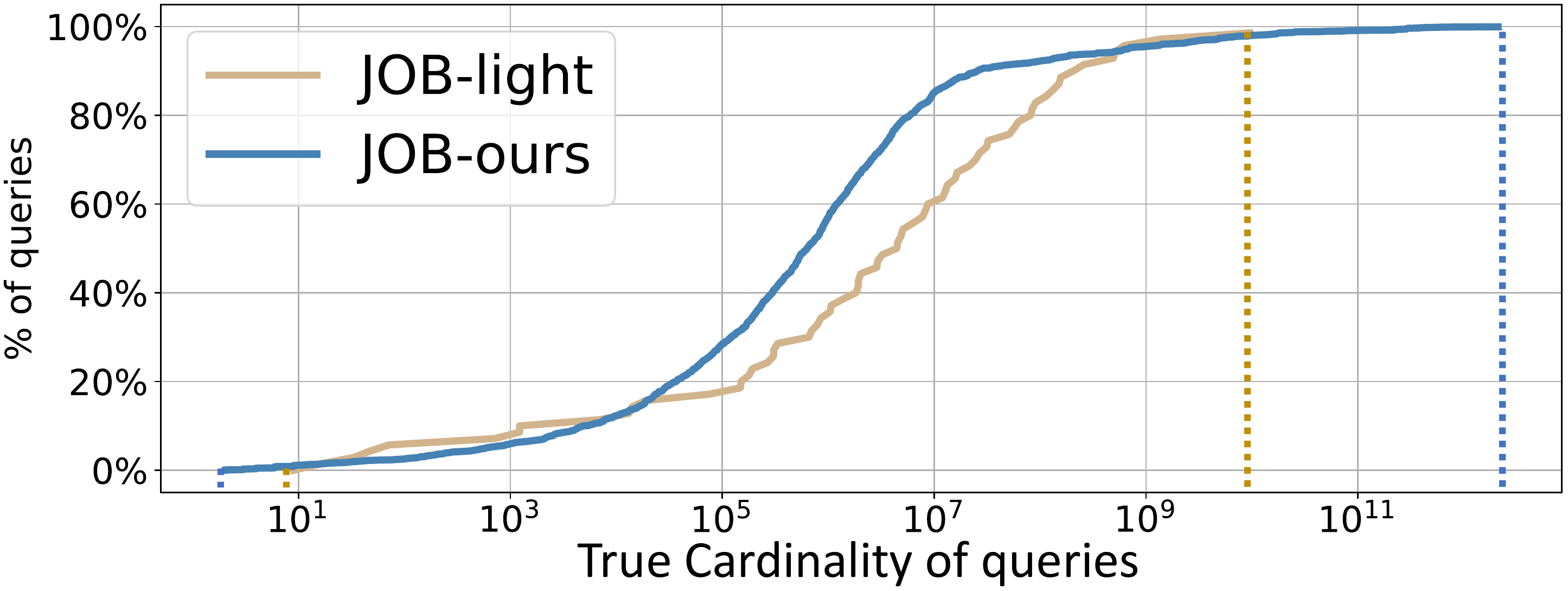}
	\caption{Cardinality distribution of workload on IMDB.}
	\label{fig: exp-card-dist}
\end{figure}

\begin{table}[t]
	\caption{Performance of \CE algorithms on \textit{JOB-light}.}
	\resizebox{0.9\columnwidth}{!}{
		\begin{tabular}{c|ccccc|c}
			\hline
			\rowcolor{mygrey}
			\bf Algorithm & \bf 50\% & \bf 90\% & \bf 95\% & \bf 99\% & \bf Max & \bf Size (KB) \\ \hline
			\textsf{Histogram} & 8.310 & 1, 386 & 6, 955 & $8 \cdot 10^5$ & $2 \cdot 10^7$ & \textbf{131} \\
			\textsf{NeuroCard} & 1.580 & 4.545 & 5.910 & 8.480  & \textbf{8.510}  & 7, 076 \\
			\textsf{BN} & 2.162 & 28.00 & 74.60 & 241.0 & 306.0 & 237 \\
			\textsf{DeepDB} & 1.250 &2.891 & 3.769 & 25.10 &31.50 & $3.7 \cdot 10^4$ \\
			\textsf{MaxDiff} & 32.31 & 5, 682 & $5 \cdot 10^4$ & $4 \cdot 10^6$ & $4 \cdot 10^7$ & $4 \cdot 10^5$ \\
			\textsf{Sample} & 2.206 & 65.80 &1, 224 & $5 \cdot 10^4$ & $1 \cdot 10^6$ &-\\ 
			\textsf{KDE} & 10.56 & 563.0 & 4, 326 & $4 \cdot 10^5$ & $8 \cdot 10^6$ & -\\
			\textsf{MSCN} & 2.750 & 19.70 & 97.60 & 622.0 & 661.0 & 3, 421\\
			\textbf{\textsf{FLAT} (Ours)} & \textbf{1.150} & \textbf{1.819} & \textbf{2.247} & \textbf{7.230} & 10.86 & 3, 430\\
			\hline
	\end{tabular}}
	\label{tab: exp-acc-qerror-multi}
\end{table}

\myskip
\noindent{\underline{\textbf{Results on \textit{JOB-ours}.}}}
On this workload, \FLAT is also the most accurate \CE method. As reported in Table~\ref{tab: exp-acc-qerror-multi-our}, we observe that: 

1) The performance of \FLAT is better than \textsf{NeuroCard} and still much better than others. At the $95\%$ quantile, \FLAT outperforms \textsf{NeuroCard}, \textsf{DeepDB} and \textsf{MSCN} by $1.4\times$,  $4.3 \times$ and $7.8 \times$, respectively. The  performance of other algorithms drops significantly on this workload. A similar observation is also reported in~\cite{yang2020neurocard}. This once again demonstrates the shortcomings of these approaches, especially for complex data and difficult queries.

2) The q-error of \FLAT on \textit{JOB-ours} is relatively larger than that on \textit{JOB-light} because \textit{JOB-ours} is a harder workload. As shown in 
Figure~\ref{fig: exp-card-dist}, the true cardinality of the tail $5\%$ queries in \textit{JOB-ours} is often less than $100$.  However, the performance of \FLAT is still reasonable since the median value is only $1.2$. 

We also examine the detailed q-errors of \FLAT and other \CE methods with different number of tables and predicates in queries. Due to space limits, we put the results in Appendix~E.3 of the technical report~\cite{fullversion}. The results show that the accuracy of our \FLAT is more stable with number of joins and predicates.

\myskip
\noindent{\underline{\textbf{Time Efficiency.}}}
Figure~\ref{fig: exp-query-multi} exhibits the average estimation latency on the two workload. Obviously, \textsf{Histogram} is still the fastest while \textsf{MaxDiff} is still the slowest. \FLAT requires around $5ms$ for each query, which is still much faster than others. It outperforms \textsf{BN} by $5 \times$, \textsf{Sample} by $12.4 \times$, \textsf{KDE} by $4.8 \times$ and \textsf{DeepDB} by $5.2 \times$. 
The training time on the IMDB dataset is given in the last column of  Table~\ref{tab: exp-acc-qerror-multi-our}. \FLAT is faster than \textsf{NeuroCard} and close to \textsf{DeepDB}.


\subsection{Effects of Updates}
\label{sec: exp-3}

\revise{We examine the performance of our incremental update method.
Specifically, for data insertion evaluation, we train the base model on a subset of IMDB data before 2004 ($80\%$ of data) and insert the rest data for updating. For data deletion, we train the base model on all data and delete the data after 1991. We compare the accuracy on the \textit{JOB-light} workload and the update time cost of our update method with two baselines: he original stale model and the new model retained on the whole data. From Table~\ref{tab: exp-multi-update}, we observe that:}

1) The accuracy of the retrained model is the highest but it requires the highest updating time. The accuracy of the non-updated model is the lowest since the data distribution changes.

2) Our update method makes a good trade-off: its accuracy is close to the retrained model but its time cost is much lower. \revise{This shows that our \FSPN model can be incrementally updated on its structure and parameters to fit the new data in terms of both insertion and deletion}. This is a clear advantage since the entire model does not need to be frequently retrained in presence of new data.

\begin{figure}[t]
	\includegraphics[width = 0.8\columnwidth]{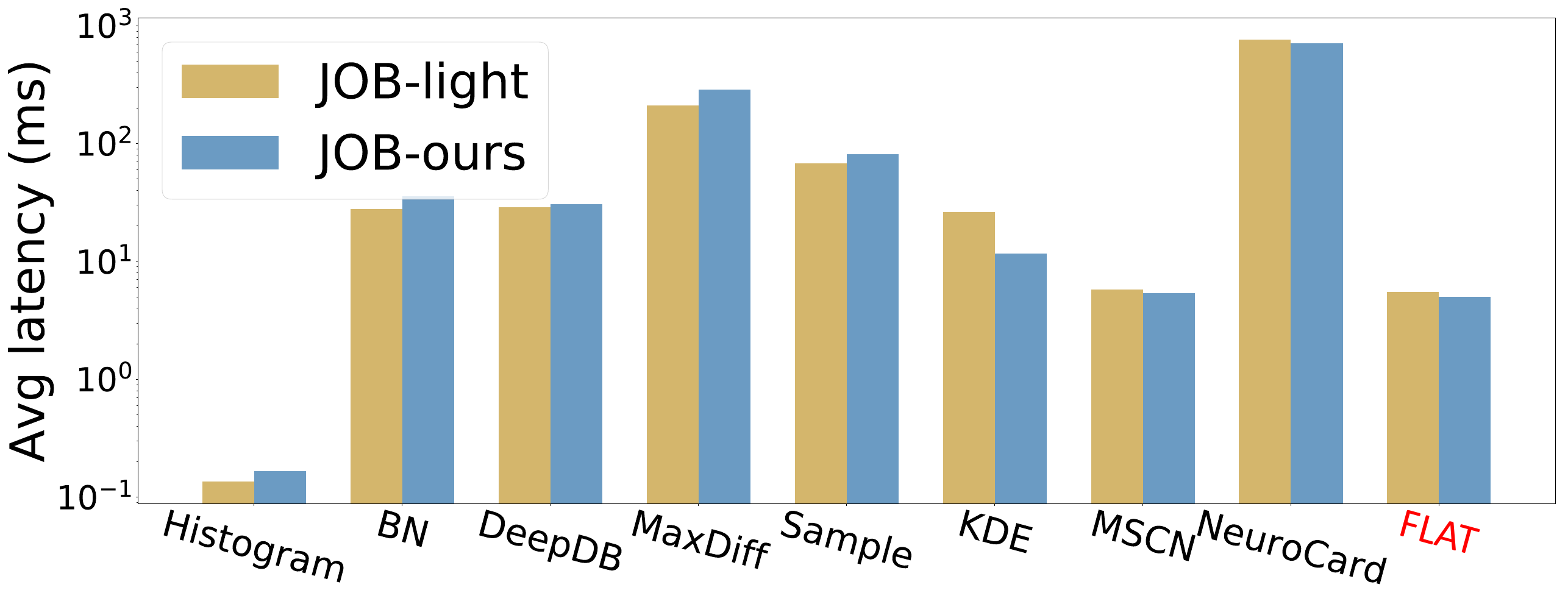}
	\caption{Estimation latency on IMDB.}
	\label{fig: exp-query-multi}
\end{figure}

\begin{table}[t]
	\caption{Performance of \CE algorithms on \textit{JOB-ours}.}
	\resizebox{0.85\columnwidth}{!}{
		\begin{tabular}{c|ccccc|c}
			\hline
			\rowcolor{mygrey}
			& & & & & & \bf Training  \\
			\rowcolor{mygrey}
			\multirow{-2}{*}{\bf Algorithm} & \multirow{-2}{*}{\bf 50\%} & \multirow{-2}{*}{\bf 90} & \multirow{-2}{*}{\bf 95\%} & \multirow{-2}{*}{\bf 99\%} & \multirow{-2}{*}{\bf Max} & \bf Time (Min)\\ 
			\hline
			\textsf{Histogram} & 15.71 & 7480 & $4\cdot10^4$ & $1\cdot10^6$ & $4\cdot10^8$ & \textbf{2.7}\\
			\textsf{NeuroCard} & 1.538 & 9.506 & 81.23 & 8012  &  $1 \cdot 10^5$ & 173\\
			\textsf{BN} & 2.213 & 25.60 & 2456 & $2\cdot10^5$ & $7\cdot10^6$ & 7.3\\
			\textsf{DeepDB} & 1.930 & 28.30 & 248.0 & $1\cdot10^4$ & $1 \cdot 10^5$ & 68\\
			\textsf{MaxDiff} & 45.50 & 8007 & $2 \cdot 10^5$ & $9 \cdot 10^6$ & $1 \cdot 10^9$ & 79\\
			\textsf{Sample} & 2.862 & 116.0 & 3635 & $3\cdot10^5$ &
			$4\cdot10^7$ & -\\
			\textsf{KDE} & 8.561 & 1230 & $1\cdot10^4$ & $9\cdot10^5$ & $2\cdot10^8$ & 15\\
			\textsf{MSCN} & 4.961 & 45.7 & 447.0 & 8576 & $1\cdot10^5$ & 1, 744\\
			\textbf{\textsf{FLAT} (Ours)} &\textbf{1.202} & \textbf{6.495} & \textbf{57.23} & \textbf{1120} & $\boldsymbol{1\cdot10^4}$ & 53\\
			\hline
	\end{tabular}}
	\label{tab: exp-acc-qerror-multi-our}
\end{table}

\begin{figure*}[t]
	\subfigure[\textbf{Query Time Excluding \CE Latency}]{\includegraphics[width = 0.355\linewidth]{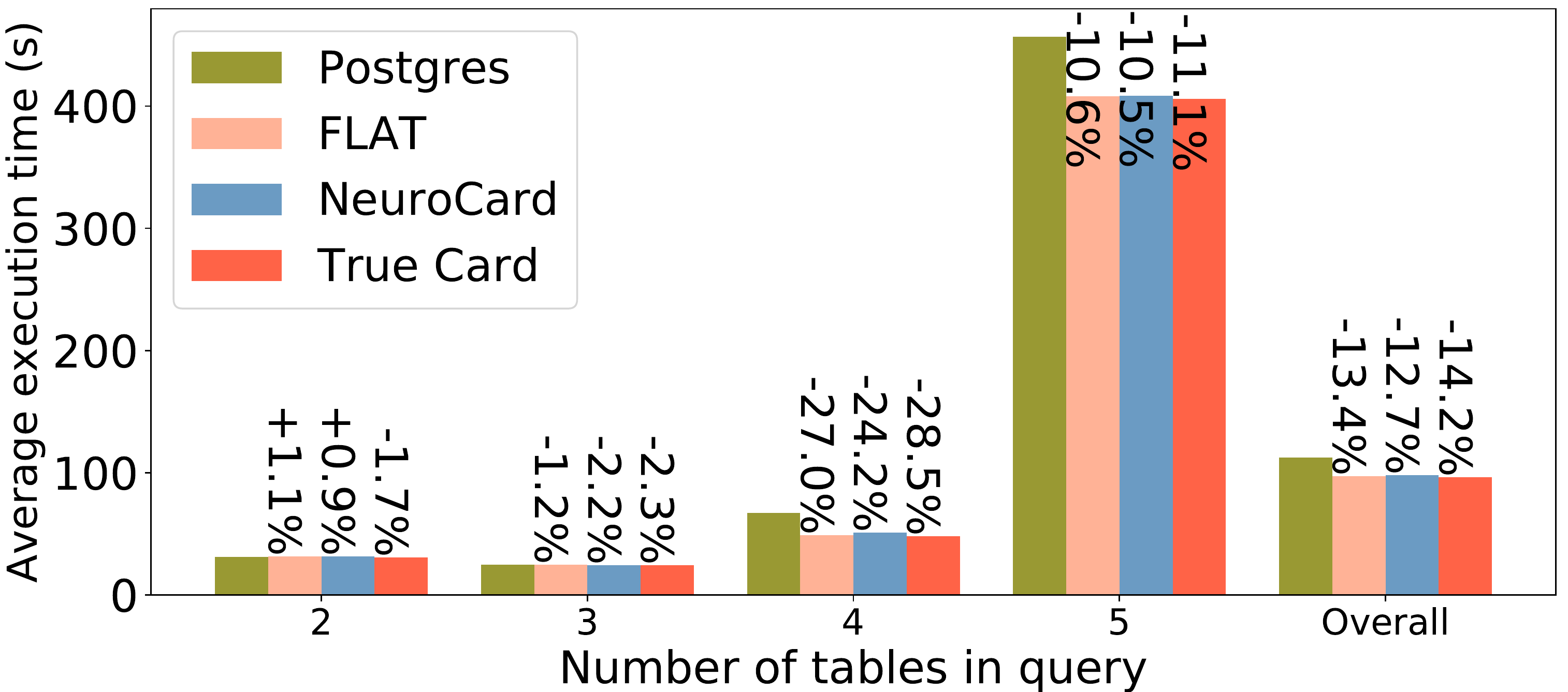}}
	\hspace{0.1\linewidth}
	\subfigure[\textbf{End-to-End Query Time}]{\includegraphics[width = 0.355\linewidth]{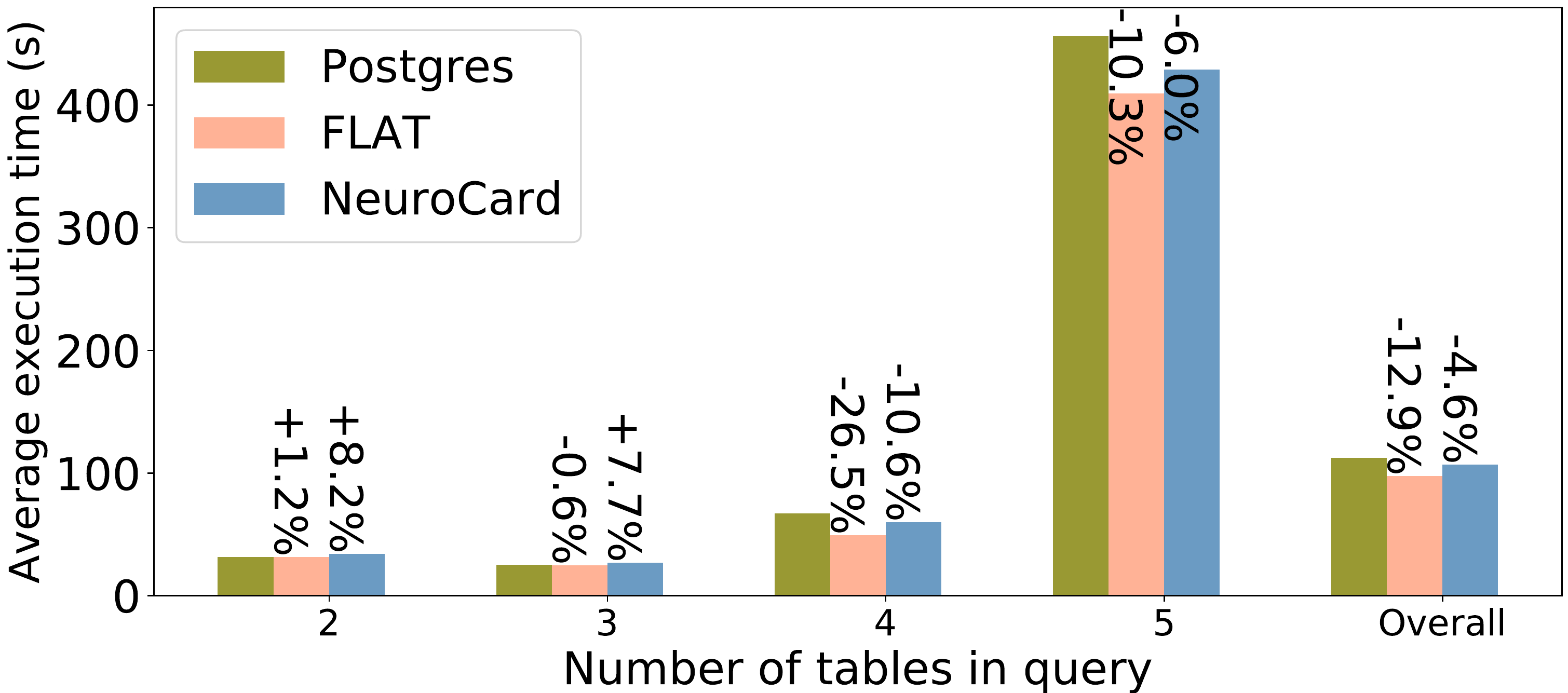}}
	\caption{Comparison of \CE algorithms integrated into Postgres.}
	\label{fig: exp-psql}
\end{figure*}

\subsection{End-to-End Evaluation on Postgres}
\label{sec: exp-4}

To examine the performance of ML-based \CE algorithms in real-world DBMS, we integrate our \FLAT and \textsf{NeuroCard} into the query optimizer of Postgres 9.6.6 to perform an end-to-end test. We do not compare with \textsf{DeepDB} since it can not support many-to-many join. However, for many star-join queries between a primary key and multiple foreign keys in the workload, the sub-queries on joining foreign keys are many-to-many joins.
Meanwhile, we add the method which uses the true cardinality of each sub-query during query optimization as the baseline.
We report the results of the \textit{JOB-light} workload on the IMDB benchmark dataset. The results on \textit{JOB-ours} are similar and put in the Appendix~E.4~\cite{fullversion}. 

We disable parallel computing in Postgres and only allow primary key indexing to minimize the impact of other factors~\cite{van2017automatic, leis2015good}.We report the total query time excluding the \CE time cost in Figure~\ref{fig: exp-psql}(a) and the end-to-end query time (including plan compiling and execution) in Figure~\ref{fig: exp-psql}(b). We observe that:

1) Accurate \CE results can help the query optimizer generate better query plans. Without considering the \CE latency, both \textsf{NeuroCard} and \textsf{FLAT} improve over Postgres by near $13\%$.
Their improvement is very close to the optimal result using true cardinality in query compiling ($14.2\%$). This verifies that the accuracy of \FLAT is sufficient to generate high-quality query plans.

2) For the end-to-end query time, the improvement of \textsf{FLAT} is more significant than \textsf{NeuroCard}. Overall, \textsf{FLAT} improves the query time by $12.9\%$ while \textsf{NeuroCard} only improves $4.6\%$. This is due to the \CE needs to do multiple times in query optimization. The latency of \textsf{NeuroCard} is much longer than \FLAT and degrades its end-to-end performance. 

3) The improvement of \FLAT becomes more significant on queries with more joins. On queries joining $4$ tables, \textsf{FLAT} improves the end-to-end query time by $26.5\%$ because the search space of the query plans grows exponentially w.r.t.~the join number.
If a query only joins $2$ or $3$ tables, its query plan is almost fixed. 
When it joins more tables, the inaccurate Postgres results may lead to a sub-optimal query plan while our \FLAT providing more accurate \CE results can find a better plan. This phenomenon has also observed and explained in~\cite{perron2019learned}. 

\begin{table}[t]
	\caption{Effects of updates on IMDB.}
	\resizebox{0.85\columnwidth}{!}{
		\color{black}
		\begin{tabular}{c|c|ccccc|c}
			\hline
			\rowcolor{mygrey}
			\bf Update & \bf Method & \bf 50\% & \bf 90\% & \bf 95\% & \bf 99\% & \bf Max & \bf Time (Min) \\ \hline
			\multirow{3}{*}{Insertion}  & Non-Updated & 1.201 & 2.297 & 3.862 & 18.93 & 47.14 & 0 \\
			& Retrained & 1.150 & 1.819 & 2.247 & 7.230 & 10.86 & 53 \\
			& \bf Our Method & 1.153 & 1.821 & 2.480 & 8.914 & 13.72 & 1.2 \\
			\hline
			\multirow{3}{*}{Deletion}  & Non-Updated & 1.218 & 2.263 & 3.905 & 15.47 & 56.21 & 0 \\
			& Retrained & 1.129 & 1.763 & 2.253 & 6.815 & 15.3 & 49 \\
			& \bf Our Method & 1.134 & 1.791 & 2.432 & 8.285 & 19.78 & 1.0
			\\
			\hline
	\end{tabular}}
	\label{tab: exp-multi-update}
\end{table}



\section{Related Work}
\label{sec: related}

We briefly review prior work on query-driven \CE methods and machine learning (ML) applied to problems in databases. The data-driven \CE methods have already been discussed in Section~2.

\myskip
\noindent{\underline{\textbf{Query-Driven \CE Methods.}}}
Initially, prior research has approached query-driven \CE by utilizing feedback of past queries to correct generated models. \revise{Representative work includes correcting and self-tuning histograms with query feedbacks~\cite{bruno2001stholes, srivastava2006isomer, fuchs2007compressed, khachatryan2015improving}}, updating statistical summaries in DBMS~\cite{stillger2001leo, wu2018towards}, and query-driven kernel-based methods~\cite{heimel2015self, kiefer2017estimating}. Later on, with the advance of deep learning, focus  shifted to learning complex mappings from ``featurized'' queries to their cardinalities. Different types of models, such as deep networks~\cite{liu2015cardinality}, tree-based regression models~\cite{dutt2019selectivity} and multi-set convolutional networks~\cite{kipf2018learned}, were applied. In general, clear drawbacks of query-driven \CE methods are as follows:
1) their performance heavily relies on the particular choice of how input queries are transformed into features;
2) they require large amounts of previously executed queries for training;
and 3) they only behave well, when  future input queries follow the same distribution as the training query samples.
Therefore, query-driven \CE methods are not flexible and generalizable enough.

\myskip
\noindent{\underline{\textbf{ML Applied in Databases.}}}
Recently, there has been a surge of interest in using ML-based methods in order to enhance the performance of database components, e.g. indexing~\cite{nathan2020learning}, data layout~\cite{kraska2018case}, query execution~\cite{park2017database} and scheduling~\cite{mao2019learning}. 
Among them, learned query optimizers are a noteworthy hot-spot.~\cite{marcus2019neo}~proposed a query plan generation model by learning embeddings for all queries.~\cite{krishnan2018learning} applied reinforcement learning to optimize the join order. We are currently trying to integrate \FLAT with these two approaches to design an end-to-end solution for query optimization in databases.

Moreover, it is worth mentioning that the proposed \FSPN model is a very \emph{general} unsupervised model, whose scope of application is not limited to \CEend. \revise{We are in the process of trying to apply to other scenarios in databases that also require modeling the joint PDF of high-dimensional data, such as
approximate group-by query processing~\cite{thirumuruganathan2020approximate}, hashing~\cite{kraska2018case} and multi-dimensional indexing~\cite{nathan2020learning}.}

\section{Conclusions}
In this paper, we propose \FLATend, an unsupervised \CE method that is simultaneously fast in probability computation, lightweight in storage cost and accurate in estimation quality. It supports queries on both single table and multi-tables. \FLAT is built on \FSPNend, a new graphical model which adaptively models the joint PDF of attributes and combines the advantages of existing \CE models. Extensive experimental results on benchmarks and the end-to-end evaluation on Postgres have demonstrated the superiority of our proposed methods.
In the future work, we believe in that \FLAT could serve as a key component in an end-to-end learned query optimizer for DBMS and the general \FSPN model can play larger roles in more database-related tasks.

\clearpage

\normalsize
\bibliographystyle{ACM-Reference-Format}
\bibliography{main}



\clearpage

\appendix

{\LARGE
\textbf{APPENDIX}
}

\setcounter{lemma}{0}
\setcounter{equation}{0}


\section{Relations between \textsf{FSPN} and Other Models}

We present here the details on how \FSPN subsumes \textsf{SPN}, as well as \textsf{BN} models. We assume that all attributes are discrete, i.e., for continuous attributes, we can discretize them by binning, and all (conditional) PDFs are stored in a tabular form.

\subsection{Transforming to \textsf{FSPN}}

We show the details on how to transform \textsf{SPN} and \textsf{BN} to the equivalent \textsf{FSPN} model.  

\smallskip
\noindent \underline{\textbf{Transforming from \textsf{SPN} to \textsf{FSPN}.}}
Given a data table $T$ with attributes $A$, if $\Pr_{\T}(A)$ could be represented by an SPN $\mathcal{S}$, we can easily construct an FSPN $\mathcal{F}$ that equally represent $\Pr_{\T}(A)$. Specifically, we disable the factorize operation in \FSPN by setting the factorization threshold to $\infty$, and then follow the same steps of $\mathcal{S}$ to construct $\mathcal{F}$. Then, the \FSPN $\mathcal{F}$ is exactly the same of $\mathcal{S}$. 

\smallskip
\noindent \underline{\textbf{Transforming from \textsf{BN} to \textsf{FSPN}.}}
Given a data table $T$ with attributes $A$, if $\Pr_{\T}(A)$ could be represented by a discrete \textsf{BN} $\mathcal{B}$, we can also build an \FSPN $\mathcal{F}$ that equally represent $\Pr_{\T}(A)$. Without ambiguity, we also use $\mathcal{B}$ to refer to its DAG structure. We present the procedures in the \textsf{BN-to-FSPN} algorithm. It takes as inputs a discrete BN $\mathcal{B}$ and the root node $N$ of $\mathcal{F}$ and outputs $F_\N$ representing the same PDF of $\mathcal{B}$. In general \textsf{BN-to-FSPN} works in a recursive manner by executing the following steps:

\indent \textcircled{1} \underline{(lines~1--3)}
If $\mathcal{B}$ contains more than one connected component
$\mathcal{B}_1, \mathcal{B}_2, \dots, \mathcal{B}_t$, the variables in each are mutually independent. Therefore, we set $N$ to be a \textsf{product} node with children $N_1, N_2, \dots, N_t$ into $\mathcal{F}$, and call \textsf{BN-to-FSPN} on $\mathcal{B}_i$ and node $N_i$ for each $i$.

\indent \textcircled{2} \underline{(lines~5--7)}
If $\mathcal{B}$ contains only one connected component, let $A_i$ be a node (variable) in $\mathcal{B}$ that has no out-neighbor. If $A_i$ also has no in-neighbor (parent) in $\mathcal{B}$, it maintains the PDF $\Pr_{\T}(A_i)$. At that time, we set $N$ to be a \textsf{uni-leaf} node representing the univariate distribution $\Pr_{\T}(A_i)$. 

\indent \textcircled{3} \underline{(lines~9--16)}
If the parent set $A_{\pa(i)}$ of $A_i$ is not empty, $A_i$ has a conditional probability table (CPT) defining $\Pr_{\T}(A_i | A_{\pa(i)}) = \Pr_{\T}(A_i | A \setminus \{A_i\})$. At this time, we set $N$ to be a \textsf{factorize} node with the left child representing $\Pr_{\T}(A \setminus \{A_i\})$ and right child $N_{\R}$ representing $\Pr_{\T}(A_i | A \setminus \{A_i\})$. For the right child $N_{\R}$, we set it to be a \textsf{split} node. For each entry $y$ of $A_{\pa(i)}$ in the CPT of $A_i$, we add a \textsf{multi-leaf} node $L_y$ of $N_{\R}$ containing all data $T_y$ in $T$ whose value on $A_{\pa(i)}$ equals $y$. On each leaf $L_y$, by the first-order Markov property of BN, $A_i$ is conditionally independent of variables $A \setminus \{ A_i \} \setminus A_{\pa(i)}$ given its parents $A_{\pa(i)}$. Therefore, we can simplify the PDF represented by $L_y$ as $\Pr_{\T}(A_i | y) = \Pr_{\T_y}(A_i)$. Therefore, $N_{\R}$ characterizes the CPT of $\Pr_{\T}(A_i | A_{\pa(i)}) = \Pr_{\T}(A_i | A \setminus \{A_i\})$. 
Later, we remove the node $A_i$ from $\mathcal{B}$ to obtain $\mathcal{B'}$, which represents the PDF $\Pr_{\T}(A \setminus \{A_i\})$. We call \textsf{BN-to-FSPN} on $\mathcal{B'}$ and node $N_{\L}$, the left child of $N$ to further model the PDF.

Finally, we obtain the FSPN $\mathcal{F}$ representing the same PDF of $\mathcal{B}$. 

\begin{figure}[!h]
	\small
	\rule{\linewidth}{1pt}
	\leftline{~~~~\textbf{Algorithm} \textsf{BN-to-FSPN$(\mathcal{B}, N)$}}
	\vspace{-1em}
	\begin{algorithmic}[1]
		\IF{$\mathcal{B}$ contains connected components $\mathcal{B}_1, \mathcal{B}_2, \dots, \mathcal{B}_t$}
		\STATE set $N$ to be a \textsf{product} node with children $N_1, N_2, \dots, N_t$
		\STATE call \textsc{BN-to-FSPN$(\mathcal{B}_i, N_i)$}
		for each $i$
		\ELSE
		\STATE let $A_i$ be a node in $\mathcal{B}$ containing no out-neighbor
		\IF{$A_i$ has no in-neighbor in $\mathcal{B}$ }
		\STATE set $N$ to be a uni-leaf node representing $\Pr_{\T}(A_i)$
		\ELSE
		\STATE  set $N$ to be a factorize node with left child $N_\L$ and right child $N_\R$
		\STATE set $N_\R$ to be a split node
		\FOR{each value $y$ of $A_{\pa(i)}$ in the CPT of $A_i$}
		\STATE add a multi-leaf node $N_y$ as child of $N_\R$
		\STATE let $T_y \gets \{ t \in T | A_{\pa(i)} \text{ of } t \text{ is } y\}$ 
		\STATE let $N_y$ represent $\Pr_{\T_y}(A_i)$
		\ENDFOR
		\STATE remove $A_i$ from $\mathcal{B}$ to obtain $\mathcal{B'}$
		\STATE call \textsf{BN-to-FSPN$(\mathcal{B}', N_\L)$}
		\ENDIF
		\ENDIF
	\end{algorithmic}
	\rule{\linewidth}{1pt}
	\vspace{-1.5em}
\end{figure}

\subsection{Proof of Lemma~1}

\indent \textbf{\textit{Lemma~1}}
\textit{
Given a table $T$ with attributes $A$, if the joint PDF $\Pr_{\T}(A)$ is represented by an \textsf{SPN} $\mathcal{S}$ or a \textsf{BN} $\mathcal{B}$ with space cost $O(M)$, then there exists an \FSPN $\mathcal{F}$ that can equivalently model $\Pr_{\T}(A)$ with no more than $O(M)$ space.
}

\begin{proof}
For the \textsf{SPN} $\mathcal{S}$ and the \FLAT $\mathcal{F}$, by Section~A.1, their structures are exactly the same. In the simplest case, if $\mathcal{F}$ represents the distribution in the same way as $\mathcal{S}$ on each leaf nodes, their space cost is the same.

For the \textsf{BN} $\mathcal{B}$ and the \FLAT $\mathcal{F}$, we analyze their space cost. The storage cost of each node $A_i$ in $\mathcal{B}$ is the number of entries in CPT of $\Pr_{\T}(A_i | A_{\pa(i)})$. The FSPN $\mathcal{F}$ represents $\Pr_{\T}(A_i)$ in step \textcircled{2} of algorithm \textsf{BN-to-FSPN} when $A_{\pa(i)}$ is empty and $\Pr_{\T}(A_i | y)$ for each value $y$ of $A_{\pa(i)}$ in step \textcircled{3}. In the simplest case, if $\mathcal{F}$ also represents the distribution in a tabular form, the storage cost is the same as $\mathcal{B}$. Therefore, the model size of $\mathcal{F}$ can not be larger than that of $\mathcal{B}$. 

Therefore, this lemma holds.
\end{proof}

\vspace{-1.5em}
\revise{\section{Detailed Updating Algorithm}}

\revise{We put the pseudocodes of our incremental updating algorithms.}

\revise{\subsection{\textsf{FLAT-Update} Algorithm}}

\revise{
The \textsf{FLAT-Update} algorithm described in Section~4.3 is used for updating \FSPN built on single table. It works in a recursive manner and traverse the original \FSPN in a top-down manner. \textsf{FLAT-Update} tries to preserve the original \FSPN structure to the maximum extent while fine-tuning its parameters for better fitting. The process to each type of nodes have been explained clearly in Section~4.3.}

\begin{figure}[t]
	\small
	\rule{\linewidth}{1pt}
	\revise{
	\leftline{~~~~{\textbf{Algorithm} \textsf{FLAT-Update$(\F, \Delta T)$}}}
	}
	\vspace{-1em}
	\revise{
	\begin{algorithmic}[1]
		\STATE let $N$ be the root node of $\mathcal{F}$
		\IF{$N$ is \textsf{factorize} node}
			\STATE call \textsf{FLAT-Update$(\F_{\N_i}, \Delta T)$} for each child $N_i$ of $N$
		\ELSIF{$N$ is \textsf{split} node}
		\FOR{each child $N_i$ of $N$}
		\STATE let $\Delta T_i$ be the set of records in the splitting condition of $N_i$
		\STATE call \textsf{FLAT-Update$(\F_{\N_i}, \Delta T_i)$}
			\ENDFOR
					\ELSIF{$N$ is \textsf{multi-leaf} node}
		\IF{the conditional indepence still holds on $T_{\N}$}
		\STATE update the parameters of the PDF by $\Delta T$
		\ELSE
		\STATE reset $N$ to be a \textsf{split} node
		\STATE call lines~28--30 of \textsf{FLAT-Offline($A_\N, C_\N, T_\N$)}
		\ENDIF
		\ELSIF{$N$ is \textsf{sum} node}
		\FOR{each child $N_i$ of $N$}
		\STATE let $\Delta T_i$ be the set of records assigned to or removed from $N_i$
		\STATE update the weights of $N_i$ accordingly
		\STATE call \textsf{FLAT-Update$(\F_{\N_i}, \Delta T_i)$}
			\ENDFOR
			\ELSIF{$N$ is \textsf{product} node}
			\IF{the indepence between attributes still holds on $T_{\N}$}
			\STATE call \textsf{FLAT-Update$(\F_{\N_i}, \Delta T)$} for each child $N_i$ of $N$
				\ELSE 
				\STATE call lines~12--19 of \textsf{FLAT-Offline($A_\N, C_\N, T_\N$)}
				\ENDIF
				\ELSIF{$N$ is \textsf{uni-leaf} node}
				\STATE update the parameters of the PDF by $\Delta T$
		\ENDIF
	\end{algorithmic} }
	\rule{\linewidth}{1pt}
\end{figure}

\vspace{-1.5em}
\revise{\subsection{\textsf{FLAT-Update-Multi} Algorithm}}

\revise{
The  \textsf{FLAT-Update-Multi} algorithm described in Technique~III of Section~5
is used to update the \FSPN models built on multi-tables.
We first identify the impact of updating $\Delta C$ onto $\mathcal{T}$, and then update the \FSPN modeling attribute columns and scattering coefficients columns, respectively. The details are clearly presented in Section~5.
}

\begin{figure}[t]
	\small
	\rule{\linewidth}{1pt}
	\revise{
	\leftline{~~~~{\textbf{Algorithm} \textsf{FLAT-Update-Multi$(\F_{\T}, \Delta C)$}}}
	}
	\vspace{-1em}
	\revise{
	\begin{algorithmic}[1]
		\IF{$\Delta C$ is inserted into table $C$}
			\STATE obtain $\Delta \mathcal{T_{+}}$ with all new records joining with $\Delta C$
			\STATE obtain $\Delta \mathcal{T_{-}}$ with all records with \textsf{null} attributes of table $C$ but can join with $\Delta C$
			\STATE obtain $\Delta \mathcal{T_{*}}$ with all original records can join with $\Delta C$
		\ELSE
			\STATE obtain $\Delta \mathcal{T_{+}}$ with all new records joining with \textsf{null} attributes of table $C$
			\STATE obtain $\Delta \mathcal{T_{-}}$ with all removed tuples joining with $\Delta C$
			\STATE obtain $\Delta \mathcal{T_{*}}$ with all original records can join with $\Delta C$
		\ENDIF
		\STATE let $N$ be the root node of $\mathcal{F}$ with left child $LC$ and right child $RC$
		\STATE call \textsf{FLAT-Update}$(\mathcal{F}_{\L \C}, \Delta \mathcal{T_{+}})$
		\STATE call \textsf{FLAT-Update}$(\mathcal{F}_{\L \C}, \Delta \mathcal{T_{-}})$
		\FOR{each \textsf{multi-leaf} node $L$ of $RC$}
			\STATE incrementally update all expected values based on the scattering columns in $\Delta \mathcal{T_{+}}$, $\Delta \mathcal{T_{-}}$ and $\Delta \mathcal{T_{*}}$ 
		\ENDFOR
		\IF{reaches the periodical updating time}
			\STATE recompute all scattering columns $S_{\T, \E}$ for all nodes $T$
			\IF{$S_{\T, \E}$ changes}
				\STATE incrementally update all expected values in all \textsf{multi-leaf} nodes
			\ENDIF
		\ENDIF
	\end{algorithmic} }
	\rule{\linewidth}{1pt}
\end{figure}

\section{Proof of Lemma~2}
\indent \textbf{\textit{Lemma~2}}
\textit{
Given a query $Q$, let $E = \{T_{1}, T_{2}, \dots, T_{d} \}$ denote all nodes in $J$ touched by $Q$. On each node $T_i$, let $S = \{ S_{\A_{1}, \B_{1}}, S_{\A_{2}, \B_{2}}, \dots, \break S_{\A_{n}, \B_{n}}\}$, where each $(A_j, B_j)$ is a distinct join such that $B_j$ is not in $Q$. Let $s = (s_1, s_2, \dots, s_n)$ where $S_{\A_{j}, \B_{j}} = s_j  \in \mathbb{N}$ for all $1 \leq i \leq n$ denote an assignment to $S$ and $\text{dlm}(s) = \prod_{j=1}^{n} \max\{s_j, 1\}$. Let}
\begin{equation}
\small
p_i \! = \frac{|\mathcal{T}_i|}{|\mathcal{E}|} \! \cdot \! 
	\sum\limits_{s, e}  \left( \Pr\nolimits_{\mathcal{T}_i}(Q_i  \wedge S \! = \! s \wedge S_{\T_{i}, \E} \! = \! e) \cdot \frac{\max\{e, 1\}}{\text{dlm}(s)} \right).
\end{equation}
\textit{Then, the cardinality of $Q$ is $|\mathcal{E}| \cdot \prod_{i = 1}^{d} p_i$.}

\begin{proof}
Given the query $Q$, let $Z$ denote all the tables touched by $Q$ and $Z_i$ be the tables touched by $Q$ in the node $T_i$. Obviously, we can obtain the cardinality of $Q$ on table $\mathcal{Z}$ as $\Card(\mathcal{Z}, Q) = \Pr_{\mathcal{Z}}(Q) \cdot |\mathcal{Z}|$.

First, we have $Z \subseteq T = \cup_{i = 1}^{d} T_i$. Let $\mathcal{E} = \mathcal{T}_1 \fullouterjoin \mathcal{T}_2 \fullouterjoin \dots \fullouterjoin \mathcal{T}_d$ denote the full outer join table over all nodes in $E$. We show how to obtain the cardinality of $Q$ on table $\mathcal{E}$.
For any single table $A \in T \setminus Z$, suppose that we have the scattering coefficient column $S_{\A, \E}$ in $\mathcal{E}$. $S_{\A, \E}$ denotes the scattering number from each record in $A$ to $\mathcal{E}$.
Let $S_{\E} = \{S_{\A_1, \E}, S_{\A_2, \E}, \dots, S_{\A_k, \E}\}$ be a collection of columns for any $A_j \in T$.
Let $s_{\E} = (s_{\A_1, \E}, s_{\A_2, \E}, \dots, s_{\A_k, \E})$, where $s_{\A_j, \E} \in \mathbb{N}$ for all $1 \leq j \leq k$, be an assignment to $S_{\E}$.
By~\cite{hilprecht2019deepdb,yang2020neurocard}, we can down-scale $\mathcal{E}$ by removing the effects of untouched tables $T \setminus Z$ to obtain the cardinality of $Q$. We have
\begin{equation}
\label{lemma2-1}
\begin{split}
\Card(\mathcal{Z}, Q) & = \Pr\nolimits_{\mathcal{Z}}(Q) \cdot |\mathcal{Z}| \\
& =
\left( \sum_{s_\E}
\frac{\Pr\nolimits_{\mathcal{E}}(Q \wedge S_{\E} = s_{\E})}{\text{dlm}(s_{\E})} \right) \cdot |\mathcal{E}|,
\end{split}
\end{equation}
where $\text{dlm}(s_{\E}) = \prod_{s \in s_{\E}} \max\{s, 1\}$.
Here, $\Pr\nolimits_{\mathcal{E}}(Q \textsc{ and } S_{\E} = s_{\E}) \cdot |\mathcal{E}|$ implies the number of records in $\mathcal{E}$ satisfying the predicate specified in $Q$ but scattered $\text{dlm}(s_{\E})$ more times by other tables in $T \setminus Z$. Therefore, we eliminate the scattering effects by dividing $\text{dlm}(s_{\E})$. We set each $s \in s_{\E}$ to $1$ when it is $0$ since records with zero scattering coefficient, i.e., having no matching, also occur once in $\mathcal{E}$. Since we do not explicitly maintain the full outer join table $\mathcal{E}$ and all columns $S_{\E}$, we need to further simplify Eq.~\eqref{lemma2-1} as follows.

Second, for each node $T_i \in E$, let $A_i = T_i \setminus Z$ denote all tables in node $T_i$ untouched by $Q$. Let $S_{\E_i} = \{S_{\A', \E} | A' \in A_i\}$ denote all the scattering columns from single table $A' \in A_i$ to $\mathcal{E}$ in $S_{\E}$. Similarly, let $s_{\E_i}$ be the assignment of $S_{\E_i}$. We could rewrite Eq.~\eqref{lemma2-1} as
\begin{equation}
\label{lemma2-3}
\small
\begin{split}
& \Card(\mathcal{Z}, Q) = \left( \sum_{s_{\E_1}, s_{\E_2}, \dots, s_{\E_d}}
\frac{\Pr\nolimits_{\mathcal{E}}(Q \wedge S_{\E_1} = s_{\E_1} \wedge \dots \wedge S_{\E_d} = s_{\E_d}}{\prod_{i=1}^{d} \text{dlm}(s_{\E_i})} \right) \cdot |\mathcal{E}| \\
& = \left( \sum_{s_{\E_1}, s_{\E_2}, \dots, s_{\E_d}}
\frac{\Pr\nolimits_{\mathcal{E}}(Q_1 \wedge S_{\E_1} = s_{\E_1} \wedge \dots \wedge Q_d \wedge S_{\E_d} = s_{\E_d}}{\prod_{i=1}^{d} \text{dlm}(s_{\E_i})} \right) \cdot |\mathcal{E}|.
\end{split}
\end{equation}
By our assumption on the join tree, the probability on different nodes are independent on their full outer join table, so we derive
\begin{equation}
\small
\label{lemma2-4}
	\begin{split}
		\Card(\mathcal{Z}, Q) 
		& = \left( \sum_{s_{\E_1}, s_{\E_2}, \dots, s_{\E_d}}
		\left( \prod_{i=1}^{d}
		\frac{\Pr\nolimits_{\mathcal{E}}(Q_i \wedge S_{\E_i} = s_{\E_i})}{\text{dlm}(s_{\E_i})} \right) \right) \cdot |\mathcal{E}| \\
		& = \left( \prod_{i=1}^{d} 
		\left( \sum_{s_{\E_i}}
		\frac{\Pr\nolimits_{\mathcal{E}}(Q_i \wedge S_{\E_i} = s_{\E_i})}{\text{dlm}(s_{\E_i})} \right) \right) \cdot |\mathcal{E}|.
	\end{split}
\end{equation}

Third, based on Eq.~\eqref{lemma2-4}, we could compute each term in the product using PDF maintained in each local node $T_i$. As stated in the Lemma, for each node $T_i$, we have scattering columns $S = \{ S_{\A_{1}, \B_{1}}, S_{\A_{2}, \B_{2}}, \dots, S_{\A_{n}, \B_{n}}\}$, where each $(A_j, B_j)$ is a distinct join where $B_j$ is not in $Q$. 
For each assignment $s = (s_1, s_2, \dots, s_n)$ of $S$, $e \in \mathbb{N}$ of $S_{\T_i, \{\E\}}$ and $s \cdot e = (s_1 \cdot e, s_2 \cdot e, \dots s_n \cdot e)$, we have 
\begin{equation}
\label{lemma2-5}
	\begin{split}
		& 
		\frac{\Pr\nolimits_{\mathcal{T}_i}(Q_i \wedge S = s \wedge S_{\T_{i}, \E } \! = \! e)}{\text{dlm}(s)} \cdot \max\{e, 1\} \cdot |\mathcal{T}_i| \\
		& = \frac{\Pr\nolimits_{\mathcal{E}}(Q_i \wedge S_{\E_i} = s \cdot e)}{\text{dlm}(s \cdot e)}
		\cdot |\mathcal{E}|.
	\end{split}
\end{equation}
We could interpret Eq.~\eqref{lemma2-5} as follows. Let $Z_i = T_i \cap Z$ be all tables being queried in node $T_i$. 
For the left hand side, we have 
\begin{equation*}
\small
\Pr\nolimits_{\mathcal{Z}_i}(Q_i) = \frac{\Pr\nolimits_{\mathcal{T}_i}(Q_i \wedge S = s)}{\text{dlm}(s)} \cdot |\mathcal{T}_i|,
\end{equation*}
which corrects the probability from $\mathcal{T}_i$ to $\mathcal{Z}_i$. Next, we should up-scale $\mathcal{Z}_i$ to the full outer join table of $Z_i \cup (E \setminus \{T_i\}) = E \setminus (T_i \setminus Z_i)$, i.e., all tables in $E$ excluding untouched tables of $Q$ in $T_i$.
Therefore, we multiply $\frac{\Pr\nolimits_{\mathcal{T}_i}(Q_i \wedge S = s \wedge S_{\T_{i}, \E}  = e) }{\text{dlm}(s)} \cdot |\mathcal{T}_i|$ by a factor of $\max\{e, 1\}$. 

\smallskip
For the right hand side, $\frac{\Pr\nolimits_{\mathcal{E}}(Q_i \wedge S_{\E_i} = s \cdot e)}{\text{dlm}(s \cdot e)}
\cdot |\mathcal{E}|$ also down-scales the probability from $\mathcal{E}$ to the full outer join table of $E \setminus (T_i \setminus Z_i)$, so it is equivalent to the left hand side.

Based on this, we have 
\begin{equation*}
\begin{split}
p_i & = \frac{|\mathcal{T}_i|}{|\mathcal{E}|} \! \cdot \! 
\sum\limits_{s, e}  \left( \Pr\nolimits_{\mathcal{T}_i}(Q_i  \wedge S \! = \! s \wedge S_{\T_{i}, \E} \! = \! e) \cdot \frac{\max\{e, 1\}}{\text{dlm}(s)} \right) \\
& = \sum_{s \cdot e} \frac{\Pr\nolimits_{\mathcal{E}}(Q_i \wedge S_{\E_i} = s \cdot e)}{\text{dlm}(s \cdot e)} \\
& = \sum_{s_{\E_i}} \frac{\Pr\nolimits_{\mathcal{E}}(Q_i \wedge S_{\E_i} = s_{\E_i})}{\text{dlm}(s_{\E_i})}.
\end{split}
\end{equation*}
Thus, the cardinality of $Q$ can be represented as
\begin{equation*}
\Card(\mathcal{Z}, Q) = |\mathcal{E}| \cdot \prod_{i=1}^{d} p_i
\end{equation*}
and the lemma holds.
\end{proof}

\revise{
\section{Sensitivity Analysis of Hyper-Parameters}}

\revise{We analyze the sensitivity of hyper-parameters of our \FLAT method.
The hyper-parameters include: 
\begin{enumerate}
\item the RDC~\cite{lopez2013randomized} threshold for deciding if two attributes are independent, denoted as $\tau_{l}$;
\item the RDC threshold for deciding if two attributes are highly correlated, denoted as $\tau_{h}$;
\item the number of intervals for $d$-way partitioning in split node;
\item the minimum amount of input data to stop further slitting the data on a node, denoted as $c$;
\item the number of bins of histograms in \textsf{uni-leaf} and \textsf{multi-leaf} nodes, denoted as $b$;
\item the number of pieces for piecewise regression in \textsf{multi-leaf} nodes, denoted as $p$;
\item the number of samples from the full outer join table to train the model, denoted as $n$.
\end{enumerate}}

\revise{All these hyper-parameters represent trade-offs between estimation accuracy and learning and/or inference efficiency. We provide the detailed analysis for each of them qualitatively as follows: 
\begin{enumerate}
\item Smaller $\tau_{l}$ would represent stricter criteria for discovering independent attributes. Thus, the model is more accurate but a large number of data-splitting operations (\textsf{sum} and \textsf{split} nodes) will be created, increasing the model size and decreasing the learning and/or inference efficiency.
\item With larger $\tau_h$, the discovered highly correlated attributes are more inter-dependent and their values are easier to model with piecewise regression, leading to more compact right branch of \textsf{factorize} nodes in \textsf{FSPN}. However, there might still exist some undiscovered highly correlated attributes that are not factorized, leading to a less compact left branch of \textsf{factorize} nodes in \textsf{FSPN}.
\item A split node with larger $d$ is more likely to break the attributes correlation. Therefore, with larger $d$, the model is potentially more compact and accurate at the same time. However, the training process would take much longer for creating each \textsf{split} node.
\item Larger $c$ can prevent \FSPN from creating a long chain of \textsf{sum} or \textsf{split} nodes, leading to more compact model with faster inference latency. However, the attributes might not be independent when a \textsf{uni-leaf} node is created, leading to less accurate estimation.
\item Intuitively, larger $b$ will help the (multi-)histograms capture the data distribution more accurately but will be more time-consuming and less space-efficient.
\item Analogously, larger $p$ will also help the \textsf{multi-leaf} nodes capture the data distribution of multiple correlated attributes more accurately but will be more time-consuming and less space-efficient. 
\item A larger sample $n$ will closely represent the actually table, prevent the model from overfitting, thus more accurate estimation. However, gathering the large number of sampled records might be time-consuming and might not be affordable in memory.
\end{enumerate}
}

\revise{
Overall, unlike DNN-based models, each hyper-parameter in \FSPN represents a trade-off, whose influence on the model is predictable. The users can easily find a set of hyper-parameters that are suitable for their datasets and affordable to the computing resources. 
}

\section{Additional Evaluation Results}

We provide some additional evaluation results in this section.

\subsection{Estimation Latency on GPUs}

In all examined \CE methods,  \textsf{MSCN} and \textsf{Naru} provide implementations specifically optimized on GPUs. We examine their query latency on a NVIDIA Tesla V100 SXM2 GPU with 64GB GPU memory. The comparison results on the dataset GAS and DMV is reported in Table~\ref{tab: exp-query-GPU}. We find that:

\begin{table}[h]
	\caption{Estimation latency on GPUs.}
	\resizebox{\columnwidth}{!}{
		\begin{tabular}{c|ccc}
			\hline
			\rowcolor{mygrey}
			& & & \bf Average \\
			\rowcolor{mygrey}
			\multirow{-2}{*}{\bf Dataset} & \multirow{-2}{*}{\bf Algorithm} & \multirow{-2}{*}{\bf Environment} & \bf Latency (ms)\\
			\hline
			\multirow{5}{*}{GAS} & \textsf{MSCN} & CPU & 3.4 \\
			& \textsf{MSCN} & GPU & 1.3 \\
			\cline{2-4}
			& \textsf{Naru} & CPU & 59.2 \\
			& \textsf{Naru} & GPU & 9.7  \\
			\cline{2-4}
			& \textsf{FLAT} & CPU & \textbf{0.5} \\
			\hline
			\multirow{5}{*}{DMV} & \textsf{MSCN} & CPU & 3.8 \\
			& \textsf{MSCN} & GPU & 1.4 \\
			\cline{2-4}
			& \textsf{Naru} & CPU & 78.7 \\
			& \textsf{Naru} & GPU & 11.5  \\
			\cline{2-4}
			& \textsf{FLAT} & CPU & \textbf{0.2} \\
			\hline
	\end{tabular}}
	\label{tab: exp-query-GPU}
\end{table}

1) Both \textsf{MSCN} and \textsf{Naru} can benefit a lot from GPU since the computation on their underlying DNNs can be significantly speed up.

2) Even without GPU acceleration, our \FLAT on CPUs still runs much faster than \textsf{MSCN} and \textsf{Naru} on GPUs. Specifically, it runs $19\times$ and $58\times$ faster than \textsf{Naru} using GPUs on GAS and DMV, respectively. This once again verifies the high estimation accuracy of our \FLAT method.

For the multi-table join query case, the \textsf{NeuroCard} algorithm provides a GPU implementation in the repository~\cite{yang2020sb}. However, they require a higher version of CUDA, which is not supported in our machine, so we do not compare with it.

\revise{\subsection{Performance Stability}}

\revise{We evaluate the stability of \FLAT using our \FSPN model in terms of the varied data criteria from four aspects: number of attributes, data distribution, attribute correlation and domain size.}

\revise{
We generate the synthetic datasets using the similar approach in a recent benchmark study~\cite{wang2020ready}. Specifically, suppose we would like to generate a table $T$ with attributes $\{A_1, A_2, \ldots,A_n\}$ and $10^6$ tuples, where $n$ denotes the number of attributes. We generate the first column for $A_1$ using a Pareto distribution\footnote{In our implementation, we use the Python library scipy.stats.pareto function}, with a value $s$ controlling the distribution skewness $s$ and a value $d$ representing the domain size. For each of the rest attribute $A_i$, we generate a column based on a previous attribute $A_j$ where $j < i$, to control the correlation $c$. Specifically, for each tuple $t = (a_1, a_2, \ldots, a_n)$ in $T$, we set $a_i$ to $a_j$ with a probability of $c$, and set $a_i$ to a random value drawn from the Pareto distribution with the probability of $1 - c$.}

\revise{
On each synthetic dataset, we generate a query workload with $1,000$ queries using the same method in Section~6.1. We report the $95\%$-quantile q-error, the average inference latency, model size and training time in Table~\ref{tab: syndata-1} and Table~\ref{tab: syndata-2}. We observe that:}

\revise{
\textbf{1. {Correlation ($s$):}} 
the varied correlation has very mild impact on \FLAT's performance. This is because \FLAT makes no independence assumption and can adaptively model the joint PDF of attributes. However, as shown in previous study~\cite{wang2020ready},  increasing the data correlation severely degrades the performance of \textsf{Histogram} and \textsf{DeepDB}.  }

\revise{
\textbf{2. {Distribution ($s$):}} 
the varied distribution skewness has slight impact on \FLAT's performance. This is because \FLAT applies (multi-)histograms to represent distributions, which are robust against distribution changes. However, as shown in previous study~\cite{wang2020ready},  increasing the Pareto distribution skewness severely degrades the performance of \textsf{Naru} and \textsf{Sample}. }

\begin{table*}[t]
	\centering
	\color{black}
	\caption{\revise{Performance stability of \FLAT w.r.t.~varied data distribution skewness and correlation.}}
	\scalebox{1.0}
	{
		\begin{tabular}{c|cccccc|cccccc}
			\hline
			\rowcolor{mygrey}
			 & & \multicolumn{4}{c}{\bf Distribution Skewness (s)} & & \multicolumn{6}{c}{\bf Attribute Correlation (c)} 
			\\
			\rowcolor{mygrey}
			\bf Data Criteria & & \multicolumn{4}{c}{$\mathbf{c=0.4, d=100, n=10}$} & & \multicolumn{6}{c}{$\mathbf{s=1.0, d=100, n=10}$}
			\\\cline{2-13}
			\rowcolor{mygrey}
			& \bf s=0 & \bf s=0.3 & \bf s=0.6 & \bf s=1.0 & \bf s=1.5 & \bf s=2.0 & \bf c=0 &\bf c=0.2 &\bf c=0.4 &\bf c=0.6 &\bf c=0.8 & \bf c=1.0 \\ \hline
			Accuracy (95\% q-error) & 1.06 & 1.15 & 1.23 & 1.76 & 2.25 & 2.11 &1.32 & 1.27 & 1.76 & 2.11 & 1.73 & 1.00 \\ \cline{1-13}
			Inference Latency (ms) & 0.6 & 0.9 & 0.6 & 0.5 &1.5 &1.7 & 0.1 & 0.7 &0.5 & 4.1 &17.8 & 0.2 \\ \cline{1-13}
			Model size (KB) & 76 & 101 & 80 & 75 &430 & 580 &9.5 &103 &75 & 1201 & 1889 & 4.7\\ \cline{1-13}
			Training time (Sec) & 91 & 93 & 127 & 142 & 240 & 253 &5.5 & 133 &244.2 & 629 & 1370 & 17.0 \\ \hline
		\end{tabular}}
	\label{tab: syndata-1}
\end{table*}

\begin{table*}[t]
	\centering
	\color{black}
	\caption{\revise{Performance stability of \FLAT w.r.t.~varied data domain size and number of attributes.}}
	\scalebox{1.0}
	{
		\begin{tabular}{c|cccccc|ccccc}
			\hline
			\rowcolor{mygrey}
			& & \multicolumn{4}{c}{\bf Domain Size (d)} & & \multicolumn{5}{c}{\bf Number of Attributes (n)} 
			\\
			\rowcolor{mygrey}
			\bf Data Criteria & & \multicolumn{4}{c}{$\mathbf{s=1.0, c=0.4, n=10}$} & & \multicolumn{5}{c}{$\mathbf{s=1.0, c=0.4, d=100}$}
			\\\cline{2-12}
			\rowcolor{mygrey}
			& \bf d=10 & \bf d=100 & \bf d=500 & \bf d=1,000 & \bf d=5,000 & \bf d=10,000 & \bf n=2 &\bf n=5 &\bf n=10 &\bf n=20 &\bf n=50 \\ \hline
			Accuracy (95\% q-error) & 1.08 &1.76 & 1.35 & 1.17 & 27.6 & 44.0 & 1.02 & 1.09 &1.76 & 12.4 & 255 \\ \cline{1-12}
			Inference Latency (ms) & 0.5 & 0.6 & 1.5 & 18.0 & 15.9 & 49.7 &0.4 &0.5 &0.5 & 3.3 & 25.9  \\ \cline{1-12}
			Model size (KB) & 16.1 &75.3 & 310 & 2701 & 1980 & 5732 & 15.0 & 49.9 &75.3 & 1780 & 6908 \\ \cline{1-12}
			Training time (Sec) & 15.5 &142 & 198 & 2670 & 1535 & 9721 &9.7 &48.6 &142 & 761 & 4017 \\ \hline
	\end{tabular}}
	\label{tab: syndata-2}
	\vspace{1em}
\end{table*}

\revise{
	\textbf{3. {Domain Size ($d$):}} the increase in the domain size degrades the performance of \textsf{FLAT}. In fact, as shown in~\cite{wang2020ready}, the performance of all \CE methods degrade with the growth of domain size. This is because increasing $d$ may rapidly increase the data complexity as there are $d^n$ possible values that a record can take. Fortunately, the performance degrades of \FLAT are still within a reasonable range. }

\revise{
\textbf{4. {Number of attributes ($n$):}} similar to domain size, the increase in the number of attributes also degrades the performance of \textsf{FLAT}. However, the performance of \textsf{FLAT} is still reasonable and affordable with tens of attributes. This is because increasing the number of attributes also increases the data complexity exponentially. In fact, the curse of dimensionality is a long-standing and common problem for almost all ML tasks. We would consider increasing the robustness of \FLAT in the future work.
}

\revise{
In summary, \FLAT is very stable to data correlation and distribution while relative robust to domain size. However, it is sensitive to the number of attributes.
}

\medskip
\medskip

\subsection{Performance with Varied Number of Joins and Predicates}

In Figure~\ref{fig: exp-multi-vary} we report the detailed q-error of \textsf{NeuroCard}, \textsf{DeepDB} and our \FLAT with different number of tables and predicates in queries. 
Clearly, when increasing the number of predicates, the q-error of  \textsf{DeepDB} significantly increases while the q-error of \textsf{NeuroCard} and \FLAT does not change too much.
When increasing join size, the performance of \textsf{DeepDB} degrades significantly while the performance of \textsf{NeuroCard} and \FLAT is affected marginally. This suggests that the joint PDF represented by \FSPN in \FLAT is more precise and robust compared to the representation via \textsf{SPN} in \textsf{DeepDB}, so its performance is more stable.
The key reasons is that \FSPN's design choices overcome the drawbacks of \textsf{SPN}. It is able to model the joint PDF of attributes with different dependency levels accordingly.

\begin{figure}[!h]
	\includegraphics[width = \columnwidth]{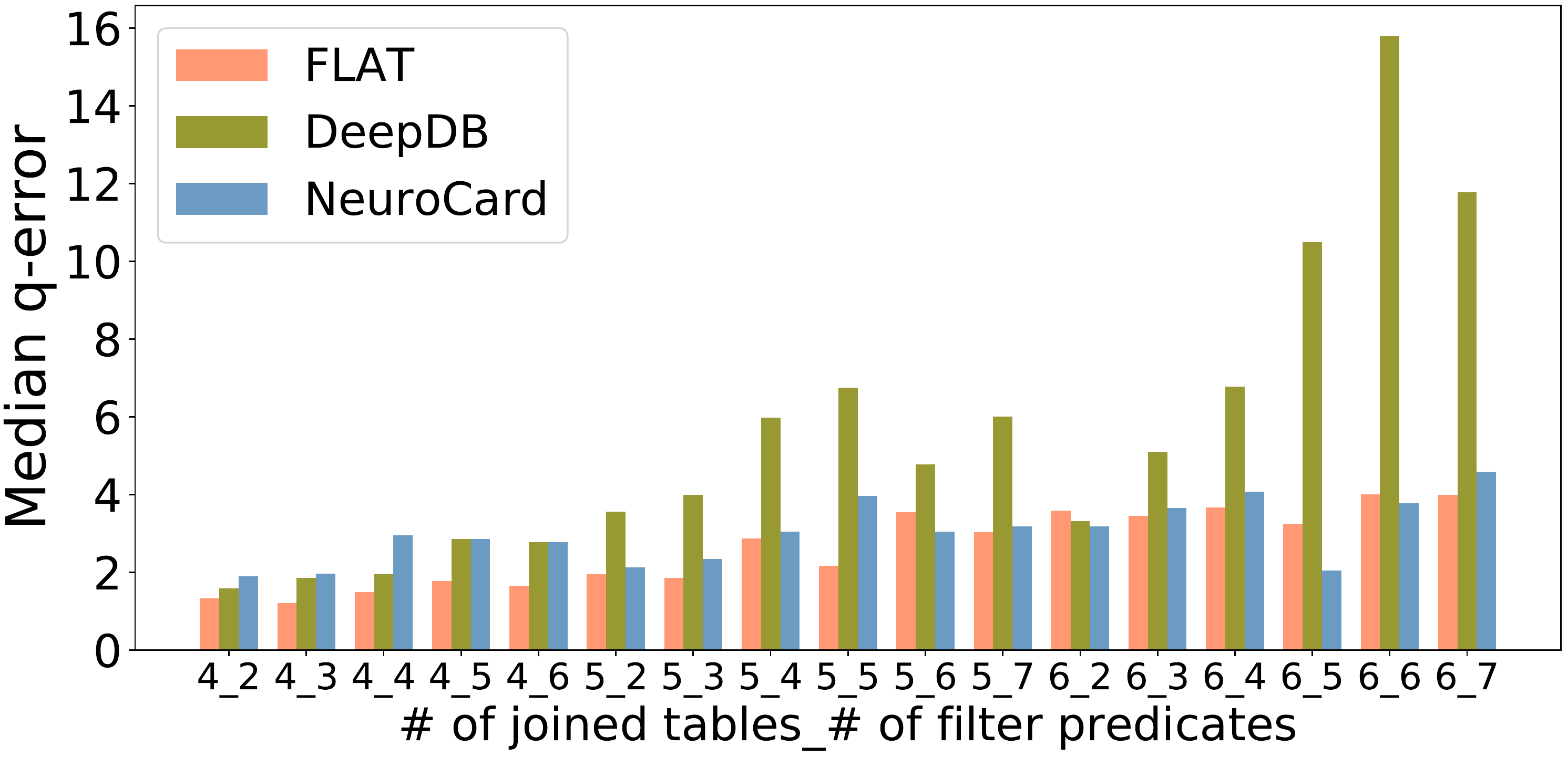}
	\caption{Q-error with different joining and predicate size.}
	\label{fig: exp-multi-vary}
\end{figure}

\subsection{End-to-End Test on \textit{JOB-ours} Workload}

We also perform the end-to-end test on Postgres using the harder workload \textit{JOB-ours}. Since \textit{JOB-ours} contains $1, 500$ queries and the execution time of some queries is extremely long, we randomly sample $50$ queries from this workload for testing.  The overall results are reported in Table~\ref{tab: exp-jobours-psql}. 

\begin{table}[t]
	\caption{End-to-end testing results on \textit{JOB-ours} workload.}
	\resizebox{\columnwidth}{!}{
		\begin{tabular}{c|ccc}
			\hline
			\rowcolor{mygrey}
			& & \bf Average Query & \\
			\rowcolor{mygrey}
			\multirow{-2}{*}{\bf Item} & \multirow{-2}{*}{\bf Algorithm} & \bf Time (Sec) & \multirow{-2}{*}{\bf Improvement} \\
			\hline
			Query Time & Postgres & 431.7 & ---\\ 
			\cline{2-4}
			Excluding & \textsf{NeuroCard} & 379.2 & $12.2\%$ \\
			\cline{2-4}
			\CE & \textsf{FLAT} &  383.3 & $11.2\%$ \\
				\cline{2-4}
			Latency & True Cardinality &  373.1 & $13.6\%$\\
			\hline
			End-to-End & Postgres & 432.3 & ---\\
				\cline{2-4}
			Query & \textsf{NeuroCard} & 405.7 & $6.2\%$ \\
				\cline{2-4}
			Time & \textsf{FLAT} &  386.9 & $10.5\%$ \\
			\hline
	\end{tabular}}
	\label{tab: exp-jobours-psql}
\end{table}

We observe similar phenomenon on the \textit{JOB-ours} workload.
Specifically, for the query time excluding the \CE query latency, the improvements of \textsf{NeuroCard} and our \FLAT are close to the method using the true cardinality.
For the end-to-end query time, the improvement of our \FLAT is more significant than \textsf{NeuroCard}.

\end{document}